\documentclass{aa}
\usepackage{graphicx}
\usepackage{rotating}
\usepackage{psfig}
\usepackage{latexsym}
\usepackage{amssymb}
\usepackage{natbib}
\usepackage{txfonts}

\newcommand{\HI}{\ion{H}{i}}
\newcommand{\HII}{\ion{H}{ii}}
\newcommand{\HeI}{\ion{He}{i}}
\newcommand{\HeII}{\ion{He}{ii}}
\newcommand{\ArII}{[\ion{Ar}{ii}]}
\newcommand{\ArIII}{[\ion{Ar}{iii}]}
\newcommand{\SIII}{[\ion{S}{iii}]}
\newcommand{\SIV}{[\ion{S}{iv}]}
\newcommand{\NeII}{[\ion{Ne}{ii}]}
\newcommand{\NeIII}{[\ion{Ne}{iii}]}

\newcommand{\OIV}{[\ion{O}{iv}]}
\newcommand{\micron}{$\mu$m}
\newcommand{\Te}{$T_{\rm e}$}
\newcommand{\den}{$n_{\rm e}$}

\def\msun{\ifmmode M_{\odot} \else $M_{\odot}$\fi}
\def\zsun{\ifmmode Z_{\odot} \else $Z_{\odot}$\fi}
\def\lsun{\ifmmode L_{\odot} \else $L_{\odot}$\fi}
\def\mup{\ifmmode M_{\rm up} \else $M_{\rm up}$\fi}
\def\mlow{\ifmmode M_{\rm low} \else $M_{\rm low}$\fi}

\begin{document}
   \title{High spatial resolution mid-infrared spectroscopy of NGC\,5253:
          The stellar content of the embedded super-star cluster\thanks{Based on
          observations obtained at the European Southern Observatory,
          La Silla, Chile (ID 70.B-0583).}}

   \author{N.\,L. Mart\'{\i}n-Hern\'{a}ndez
          \inst{1}
          \and
          D. Schaerer\inst{1,2}
          \and
          M. Sauvage\inst{3}
          }

   \offprints{N.L.\,Mart\'{\i}n-Hern\'{a}ndez,\\
             \email{Leticia.Martin@obs.unige.ch}}

   \institute{
              Observatoire de Gen\`{e}ve, 51 Chemin des Maillettes,
              CH-1290 Sauverny, Switzerland
        \and
        Laboratoire Astrophysique de Toulouse-Tarbes (UMR 5572),
        Observatoire Midi-Pyr\'en\'ees,
        14 Avenue E. Belin, F-31400 Toulouse, France
         \and
              CEA/DSM/DAPNIA/SAp, CE Saclay, 91191 Gif sur Yvette
              Cedex, France
            }

   \date{}

\titlerunning{Mid-IR spectroscopy of the embedded SSC in NGC\,5253}
\authorrunning{Mart\'{\i}n-Hern\'{a}ndez et al.}

   \abstract{
We present the $N$-band (8--13~\micron) spectrum of the
hidden compact radio super-star cluster in NGC\,5253, C2, obtained
with TIMMI2 on the ESO 3.6\,m telescope. The spectrum is characterised
by a rising continuum due to warm dust, a silicate absorption and a
strong \SIV\ line at 10.5~\micron. Weaker lines of \ArIII\ at
9.0~\micron\ and \NeII\ at 12.8~\micron\ are also present.  The
continuum can be modeled by an optically thick emission from hot
($T_{\rm d}=253 \pm 1$~K) dust emission extinguished by a cold
foreground dust screen and a silicate absorption feature with $A_{\rm
sil} = 0.73 \pm 0.05$ mag. We show how the spatial scale of the
observations greatly determine the mid-IR appearance of NGC\,5253 and
the important implications that this has on the interpretation of line
fluxes in terms of the properties (age, IMF, etc.) of the embedded
cluster.

We have modeled the observed line fluxes towards C2 using
photoionisation models with the most recent spectral energy
distributions available to describe the integrated properties of the
stellar cluster.  The detailed dependence of the mid-IR lines on
parameters such as the cluster age, upper mass cutoff and power law
index of the IMF, as well as the local abundance, the presence of
internal dust and the density structure is largely discussed.  Strong
constraints on the geometry based on high spatial resolution
observations at different wavelengths -- near-IR (HST and Keck),
mid-IR (TIMMI2) and radio (VLA) -- allows us to restrain the
ionisation parameters to values $\log U \ge -0.5$~dex.  This
constraint on $U$ lead to two possible solutions for the age and upper
mass cutoff of C2: {\em 1)} a young ($< 4$~Myr) cluster with a
"non-standard" IMF with a low upper mass cutoff \mup $< 50$\,\msun,
and {\em 2)} a cluster of $\sim 5-6$~Myr with a standard high upper
mass cutoff (\mup $\sim 100$\,\msun).  A young age ($<$ 4 Myr) would
agree with the lack of supernovae signatures in C2 and in case of
being confirmed, would be the first indication for a ``non-standard'',
low upper mass cutoff of the IMF for an individual massive cluster. An
older age of $\sim$ 5--6~Myr would imply that it is possible to
``contain'' and hide such a compact cluster for a longer time that
what it is generally thought.  Arguments in favour and against these
two scenarios are presented.

The origin of the \OIV\ 25.9~\micron\ emission measured by ISO and the
possible presence of an intermediate mass black hole inside C2 are
also addressed.
 }

   \maketitle
%

\section{Introduction}
Near to mid-IR observations suffering from less or little extinction
are thought to provide more accurate spectral diagnostics to quantify
massive star formation and to distinguish between stellar (starburst)
and other (AGN) activity \citep[cf.][]{genzel00}.
With the near routinely access to mid-IR observations and thanks to
their increased sensitivity, diagnostics in this spectral domain have
been measured for a large number of Galactic and extra-galactic
objects with the {\it Infrared Space Observatory} (ISO).  Furthermore,
observations with the {\it Spitzer Space Telescope} have just begun.

Despite the important advantage of strongly reduced extinction, mid-IR
satellite observations are potentially hampered by limitations due to
their large aperture, especially for extra-galactic targets.  As high
spatial resolution mid-IR observations show, starburst galaxies and
related objects can exhibit complex structures, numerous ``knots''
etc.  both in the continuum as in emission lines
\citep[e.g.][]{boker98,soifer01,soifer02} on scales much smaller than
the spectroscopic apertures (e.g.\ typically 280--380 arcsec$^2$ for
ISO/SWS).  Furthermore, different diagnostics (e.g.\ PAH and continuum
emission, high and low excitation fine structure lines) may originate
from different spatial regions.  In consequence, it is by no means
clear if and to what extent such spatially integrated or ``global''
spectra can be used for various diagnostics purposes.

For example, the reason for a relatively low average excitation in
starburst galaxies as measured by the fine structure line ratio
\NeIII/\NeII\ \citep[cf.][]{thornley00} is not completely clear and
has led to somewhat conflicting or inconsistent interpretations.
\cite{thornley00} argue e.g.\ that the low \NeIII/\NeII\ ratios are
due to short burst timescales and aging effects allowing, however, for
a ``normal'' Initial Mass Function (IMF) including massive stars up to
$\approx$ 50--100 \msun.
In contrast, \cite{rigby04} argue for a deficiency of stars with $M
\ga 40 \msun$ or suggest that massive stars or star clusters in
massive galaxies spend a significant fraction of their lifetime hidden
even at near to mid-IR wavelengths.
\cite{martin:metal}, on the other hand, have pointed out a dependence
of the observed \NeIII/\NeII\ ratio of starbursts on metallicity and
argue that a proper treatment of the metallicity effects on stellar
tracks, atmospheres and nebular models must be taken into account in
the comparisons.
Finally one may wonder how reliable nebular models assuming a single
``average'' ionisation parameter are and to which extent the low
\NeIII/\NeII\ ratio is due to a non-negligible contribution of \NeII\
from a diffuse low excitation medium \citep[cf.][]{schaerer03}.
Although  \cite{thornley00} estimate a diffuse contribution to be
insignificant, this remains to be demonstrated observationally.

In short, given the complexity of modelling and interpreting large
aperture fine structure line emission and possibly other mid-IR
diagnostics, there appears to be an important need for clarification
based on simple objects representing, ideally, a single age and
metallicity population with a well constrained geometry. The latter,
in particular, is of prime importance to constrain the ionisation
parameter of the nebular model and has generally not been treated with
enough care.

With these aims in mind we have recently started gathering
ground-based high spatial resolution mid-IR observations (imaging and
spectroscopy) of several nearby starbursts.  Among them is the well
studied bright starburst galaxy NGC\,5253, a nearby Magellanic type
irregular, observed at basically all wavelengths from X-ray to radio.
Although this galaxy has a lower metallicity and higher excitation
than the typical starbursts observed with ISO
\citep[cf.][]{thornley00,verma03}, the observations and analysis
present here can provide interesting insight on the difficulties
related to low spatial resolution spectroscopic observations and on
the problems discussed above.

The galaxy NGC\,5253 is also of prime interest for studies related to
super star clusters (hereafter SSCs), as it is one of the first
starbursts where a very compact and hidden SSC was found from high
spatial resolution radio observations \citep{turner98}.  Such
optically thick bremsstrahlung radio sources represent most likely the
earliest phases of SSCs observed so far
\citep[e.g.][]{kobulnicky99,vacca02}. As such, they provide invaluable
insight on the properties of SSCs, their formation and evolution.
Furthermore, studying the youngest star forming regions yields in
principle the most accurate constraints on the true upper limit of the
IMF, since least affected by effects such as mass segregation,
evaporation, and by the short finite lifetime of the most massive
stars.

Interestingly, the compact radio SSC in NGC 5253, often also referred
to as the radio ``supernebula'', also stands out by its large
contribution to the total integrated flux of the galaxy at various
wavelength.  For instance, \cite{gorjian01} estimate that at least
15\% of the bolometric luminosity of NGC 5253 is due to this SSC.
\cite{vanzi04} find that it dominates the total galaxy emission at
wavelength longer than 2--3 \micron.
Recently, HST/NICMOS observations have permitted resolving the region
close to the supernebula revealing the presence of a nuclear double
star cluster separated by 0\farcs3--0\farcs4 \citep{alonso04}. The
eastern near-IR cluster (C1) is identified with a young optical
cluster, whereas the western star cluster (C2) coincides with the
radio supernebula.  Keck observations of NGC\,5253~C2
\citep{gorjian01} at 11.7~\micron\ reveal an unresolved source with a
diameter of 0\farcs58 and a total flux of 2.2~Jy.  Radio observations
at mm and cm wavelengths \citep{turner00,turner04} show a slightly
elongated source of about $\sim$\,0\farcs1$\times$0\farcs05.  As we
will see below, these high spatial resolution observations combined
with our mid-IR spectroscopy provide, by means of detailed
photoionisation models, a unique opportunity to study in detail the
properties of this extreme hidden massive star cluster.

The core of the paper is structured as follows.  Our observations are
described in Sect.\ \ref{sect:observations}.  Immediate results from
our spectra are discussed in Sect.\ \ref{sect:results}.  Detailed
photoionisation models for the compact SSC C2 are presented in Sect.\
\ref{sect:modeling}.  Our principal results on the age, initial mass
function and other properties of the SSC, as well as various
implications are discussed in Sect.\ \ref{sect:discussion}.  The main
conclusions of the paper are summarised in Sect.\
\ref{sect:conclusions}.

\section{Observations and data reduction}
\label{sect:observations}

Our new infrared data on NGC\,5253 were obtained as part of a
programme with the Thermal Infrared MultiMode Instrument (TIMMI2) on
the ESO 3.6\,m telescope (La Silla Observatory, Chile) to observe
young starburst galaxies.

The $N$-band spectrum of NGC\,5253~C2 was obtained on 2003 March 21.  We
used the 10~\micron\ low-resolution grism which ranges from 7.5 to
13.9~\micron\ and has a spectral resolving power
$\lambda/\Delta\lambda \sim 160$. The slit width used was
1\farcs2$\times$70\arcsec, with a pixel scale of 0\farcs45.  The slit
was positioned across the bright infrared supernebula C2 and in the north-south
direction. In order to correct for background emission from the sky,
the observations were performed using
a standard chopping/nodding technique along the slit (where the object
is observed at two different positions on the slit) with an
amplitude of 10\arcsec\ in the north-south direction.
The effective exposure time (on-source) was 32 min. The
observation was performed at an airmass of $\sim 1.2$. The star
HD\,123139 was observed right before and after C2 and
served as both
telluric and flux standard star. This is a primary  ISO calibration
standard star and is described in detail at the TIMMI2
webpage\footnote{www.ls.eso.org/lasilla/sciops/timmi}. The synthetic
calibrated spectrum for this standard star is  given by
\citet{cohen99}.

   \begin{figure*}[!ht]
   \centering \includegraphics[width=18cm]{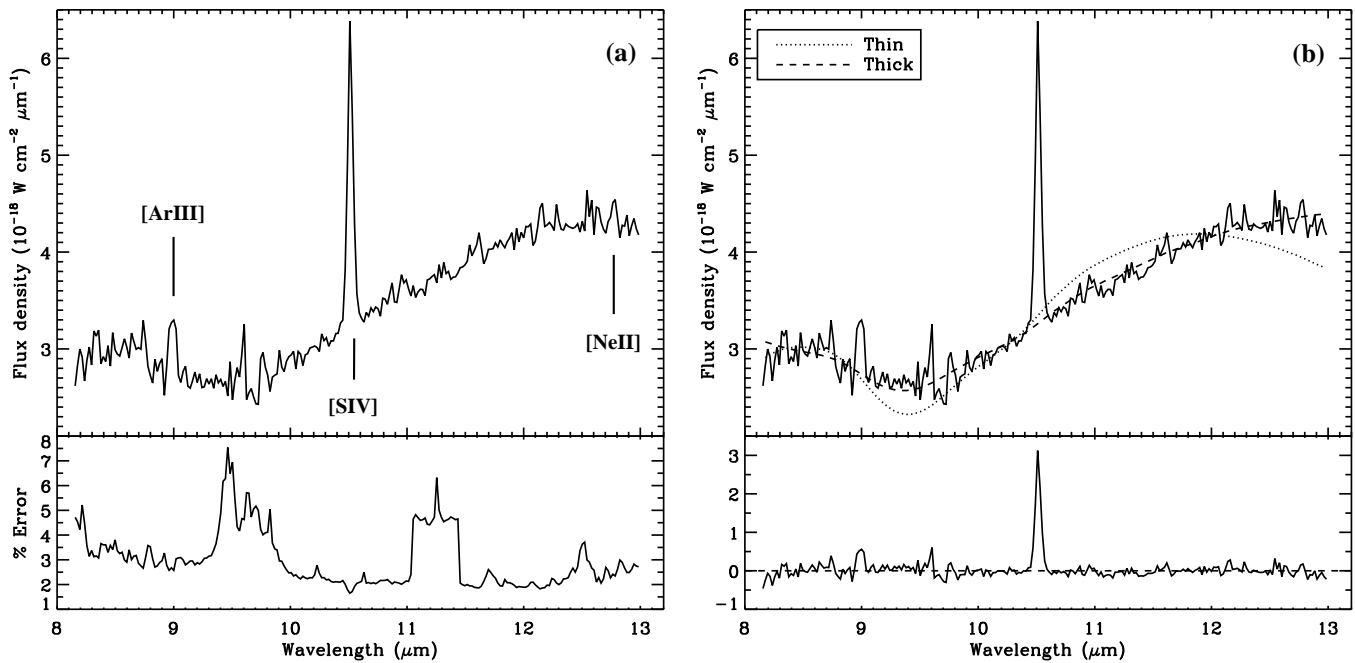}
   \caption{{\bf (a)} Top panel shows the $N$-band spectrum of
   the embedded cluster NGC\,5253~C2.
   The spectrum is characterised by a rising dust
   continuum and by the atomic fine-structure lines of \ArIII\ at 9.0
   \micron, \SIV\ at 10.5 \micron\ and \NeII\ at 12.8 \micron.
   Bottom panel shows the statistical uncertainty at each
   wavelength.
   {\bf (b)} Top panel shows the best optically thin and
   optically thick models for the
   dust continuum emission (see Sect.~\ref{sect:extinction}).
   Bottom panel shows the continuum-subtracted spectrum when applying
   the optically thick model for the continuum.}
         \label{fig:spectrum}
   \end{figure*}

The data processing included the removal of bad frames and the coaddition
of all chopping and nodding pairs. This left us with one single
image with one positive and two negative long-slit spectra.
These were combined with a simple shift-and-add
procedure which increased significantly the signal-to-noise.
The spectra of the target and standard star were
extracted using the optimal extraction procedure developed by
\cite{horne86}, ideal for unresolved point sources.
This procedure applies non-uniform pixel weights in the
extraction sum in order to reduce the statistical noise in the extracted
spectrum to a minimum while preserving its photometric accuracy.
Slit losses are negligible, since the full-width-at-half-maximum (FWHM)
along the spatial direction for both target and standard star is
1\farcs1, smaller than the slit width (1\farcs2).
The calibration of the spectroscopic data included {\em 1)} the  removal of the
telluric features, which was done by dividing by  the spectrum of
the standard star, {\em 2)} the removal of the spectral features of the
standard star and {\em 3)} the absolute flux calibration. These last two
steps were conjointly done by multiplying by the synthetic spectrum of the
standard star.  Wavelength calibration is straight forward since a
table with the pixel-to-wavelength correspondence is provided by the
TIMMI2 webpage.

\section{Results}
\label{sect:results}

\subsection{The N-band spectrum}

Figure~\ref{fig:spectrum}\,a shows the spectrum of the embedded cluster
NGC\,5253~C2. Only the valid range between 8.15 and 13~\micron\ is shown.
The spectrum is characterised by a rising continuum due to warm dust, a
silicate absorption around 9.7~\micron\ and the fine-structure
lines of \ArIII\ at  9.0 \micron, \SIV\ at 10.5 \micron\ and \NeII\ at
12.8 \micron. The \NeII\ line is very weak.
The presence of silicate absorption contradicts the best
fitting model of the nuclear SED by \cite{vanzi04}, which does not
require a silicate component.
The ISO/SWS spectrum of NGC\,5253 by \cite{crowther99} shows no
evidence of absorption at the location of the silicate band. However,
the silicate absorption might be hampered by the large ISO aperture
($\sim$ 14\arcsec$\times$20\arcsec) and by the low signal-to-noise of the
spectrum. A direct comparison with the ISO/SWS spectrum also shows that the
continuum flux level of NGC\,5253~C2 is about 3 times lower that that
observed by the large ISO aperture.

The spectrum does not display PAH emission features. This is confirmed by
the $L$-band spectrum of the nuclear region of NGC\,5253 obtained by
\cite{alonso04}, which shows the non-presence of the 3.3~\micron\ PAH feature.
The ISO spectrum integrated over the whole galaxy \citep{crowther99}
shows no signature of PAH emission features neither.
Most massive star-forming galaxies show emission bands at 3.3, 6.2, 7.7, 8.6
and 11.2~\micron, generally attributed to emission by
Polycyclic Aromatic Hydrocarbon (PAH) molecules
\citep[e.g.][]{allamandola89}. Hence, the non-detection of PAH features
towards NGC\,5253~C2 makes it look more like an AGN than a star forming region
\cite[e.g.][]{genzel00,siebenmorgen04}.
However, it is shown from ISO observations that low metallicity dwarf galaxies
such as NGC\,5253 tend to show weak of no mid-IR PAH emission features
\citep[e.g.][]{thuan99,madden02}. Moreover, PAH features are not present in
the mid-IR spectra of highly embedded massive star formation sites
\citep[e.g.][]{peeters04}.
The effects of low metallicity and a
harsh environment in NGC\,5253~C2 probably suppress the formation of PAH
emission features.

\begin{table*}[!ht]
\caption{Line fluxes of NGC\,5253 and its embedded cluster C2.}
  \label{table:fluxes}
  \begin{center}
    \leavevmode
    \begin{tabular}[h]{lcr@{\,}c@{\,}lr@{\,}c@{\,}lcc}
      \hline \hline \\[-7pt]
    \multicolumn{1}{c}{Line} &
    \multicolumn{1}{c}{$\lambda$} &
    \multicolumn{3}{c}{Line flux$^a$} &
    \multicolumn{3}{c}{Line flux$^b$} &
    \multicolumn{1}{c}{$A_\lambda$} &
    \multicolumn{1}{c}{Aperture} \\
    \multicolumn{1}{c}{ } &
    \multicolumn{1}{c}{(\micron)} &
    \multicolumn{6}{c}{(10$^{-20}$ W\,cm$^{-2}$)} &
    \multicolumn{1}{c}{(mag)} &
    \multicolumn{1}{c}{(\arcsec)}
    \\[5pt] \hline \\[-7pt]

\multicolumn{9}{l}{This work:}\\[5pt]

\ArIII & 9.0  &  4.4 &$\pm$& 1.0 &  7.6 &$\pm$& 1.8 & 0.60 $\pm$0.04 &1.2\\
\SIV   & 10.5 & 18.0 &$\pm$& 0.7 & 29.9 &$\pm$& 1.6 & 0.55 $\pm$0.04 &1.2\\
\NeII  & 12.8 &  1.4 &$\pm$& 0.5 &  1.7 &$\pm$& 0.6 & 0.23 $\pm$0.02 &1.2\\[5pt]
\hline \\[-7pt]

\multicolumn{9}{l}{\citet{turner03}:}\\[5pt]

Br$\gamma$ & 2.17 & 0.4 &$\pm$& 0.1 & 2.2 &$\pm$& 1.4 & 1.9 $\pm$ 0.6 &0.579\\
Br$\alpha$ & 4.05 & 3.4 &$\pm$& 1.0 & 6.2 &$\pm$& 2.2 & 0.7 $\pm$ 0.2 &0.579\\[5pt]
\hline \\[-7pt]

\multicolumn{9}{l}{\citet{lumsden94}:}\\[5pt]

\HeI & 2.06 & 0.240&$\pm$&0.004 & 1.6&$\pm$&1.0 & 2.1$\pm$0.7 &3\\[5pt]

\hline \\[-7pt]

\multicolumn{9}{l}{Selected mid-IR line fluxes measured by ISO/SWS \citep{verma03}:}\\[5pt]

\ArII  & 7.0  &  2.3 &$\pm$& 0.2 & 2.6  &$\pm$& 0.2 &
0.13$\pm$0.01 & $14\times20$\\

\ArIII & 9.0  &  5.9 &$\pm$& 0.5 &10.2  &$\pm$& 0.9 &
0.60$\pm$0.04 & $14\times20$\\

\SIV   &10.5  &   34 &$\pm$& 3   &   56 &$\pm$& 5   &
0.55$\pm$0.04 & $14\times20$\\

\NeII  &12.8  &    8 &$\pm$& 1   &   10 &$\pm$& 1   &
0.23$\pm$0.02 & $14\times27$\\


\NeIII & 15.5 & 28.9 &$\pm$& 4.6 &  34  &$\pm$& 5   &
0.19$\pm$0.01 & $14\times27$\\

\SIII  & 18.7 &  15.9&$\pm$& 2.5 &  20  &$\pm$& 3   &
0.25$\pm$0.02 &  $14\times27$\\

\OIV   & 25.9 & 0.9 &$\pm$& 0.1  & 1.0  &$\pm$& 0.1     &
0.14$\pm$0.01 &  $14\times27$\\[5pt]\hline

    \end{tabular}
  \end{center}
{\small
$^a$ Before extinction correction.
$^b$ After extinction correction.
}
\end{table*}

\subsection{Line fluxes observed at different spatial resolutions}
\label{sect:lines}

Line fluxes in the $N$-band spectrum were measured by fitting a
Gaussian and are listed in Table~\ref{table:fluxes}.  We have compared
these line fluxes with IR measurements for  NGC\,5253 from the
literature.

\citet{beck96} observed the nuclear region of the galaxy with a
resolution of 1\farcs6 and obtained fluxes for the \ArIII, \SIV\ and
\NeII\ lines of, respectively,  $3.6 \times 10^{-20}$, $39 \times
10^{-20}$ and $< 1 \times 10^{-20}$ W\,cm$^{-2}$. The  \ArIII\
line flux is comparable to the one we have measured with TIMMI2, while the
line flux of \SIV\ is about 2 times higher. The reason of this discrepancy
is not clear to us.

The central region of NGC\,5253 has also been observed by the SWS
on board ISO, resulting in a
spectrum ranging from 2.3 to  45~\micron\ \citep{crowther99}. The
ISO/SWS apertures are approximately $14\arcsec \times 20\arcsec$ up to
12~\micron, and  $14\arcsec \times 27\arcsec$ between 12 and
19.6~\micron. The line fluxes measured by ISO/SWS are  $(5.9 \pm 0.5)
\times 10^{-20}$, $(34 \pm 3) \times 10^{-20}$ and $(8 \pm 1) \times
10^{-20}$ W\,cm$^{-2}$, respectively for the \ArIII, \SIV\ and  \NeII\
lines \citep[][ see also Table~\ref{table:fluxes}]{verma03},
where we have adopted as uncertainties the
typical ISO/SWS $1\sigma$ absolute flux accuracy
\citep{peeters:catalogue}. Line ratios between the ISO/SWS and TIMMI2
fluxes are $1.3 \pm 0.3$, $1.9 \pm 0.1$ and $5.7 \pm 2.0$.
Ar$^{++}$, S$^{3+}$ and Ne$^+$ have ionisation potentials of 27.6,
34.8 and 21.6 eV, respectively. Hence, in view of the comparison between the
line fluxes measured towards NGC\,5253~C2 by TIMMI2 and those measured
by the large
ISO aperture, we see that a large fraction of the emission
from the high-excitation lines (\ArIII\ and \SIV) in the
central starburst region of NGC\,5253 comes from the compact
and embedded \HII\ region associated to NGC\,5253~C2. However, only
$\sim$20\% of the \NeII\
emission comes from this \HII\ region.
Other sources such as the remaining observed clusters and the diffuse ISM
must contribute to the \NeII\ line flux measured by ISO/SWS.

Ratios of fine-structure lines are commonly used in a diversity
of diagnostics. For instance,
the ratio of the high-to-low-excitation emission lines
[\ion{O}{iv}]\,25.9/\NeII\,12.8~\micron\ is used, in combination with
the 7.7 \micron\ PAH line-to-continuum ratio, to separate AGN
dominated and star formation dominated nuclei \citep[cf.][]{genzel98}). 
As well,
ratios such as \NeIII\,15.5/\NeII\,12.8~\micron\ are used to infer the
age of the starburst \citep[e.g][]{thornley00}.
The presence of IR-bright hidden clusters such as the one observed in
NGC\,5253 and their contribution to the global atomic line emission
might thus obfuscate diagnostic diagrams used to determine the
ultimate physical process  powering galactic nuclei and studies of the IMF.

We also compare the Br$\alpha$ line fluxes measured by
\cite{turner03} using NIRSPEC on the Keck Telescope--
with an aperture of 0.579\arcsec\ centred on
NGC\,5253~C2 -- and by ISO/SWS. These fluxes are, respectively,
$(3.4 \pm 1.0)\times10^{-20}$ (see Table~\ref{table:fluxes}) and
$(4.5 \pm 0.4) \times 10^{-20}$~W\,cm$^{-2}$ \citep{verma03}.
Both line fluxes are in
good agreement, indicating that practically all the Br$\alpha$ emission is
produced
by NGC\,5253~C2.

\subsection{Extinction}
\label{sect:extinction}

\subsubsection{Extinction in the near-IR}
\label{sect:ext:nearIR}

\cite{turner03} observed the Brackett $\gamma$ (2.17~\micron) and
$\alpha$ (4.05~\micron) recombination lines of hydrogen towards NGC\,5253~C2
using NIRSPEC on the Keck Telescope.
The observed and extinction-corrected line fluxes are listed in
Table~\ref{table:fluxes}. The lines were corrected for extinction
considering that the intrinsic $F_{\alpha}/F_{\gamma}$ flux ratio is
2.8 (for an electron temperature $T_{\rm e}=12\,000$~K;
\citeauthor{hummer87} \citeyear{hummer87})
and an extinction law $A_{\lambda} \propto \lambda^{-1.7}$
\citep{mathis90}. The value of the electron temperature is determined
for visible \HII\ regions in NGC\,5253
\citep{campbell86,walsh89,kobulnicky97}. We have recomputed the
extinction values obtained by \cite{turner03} in order to include a detailed
error analysis. The extinctions with their respective uncertainties are:
$A_{\gamma} = 1.9 \pm 0.6$~mag, $A_{\alpha} = 0.7 \pm 0.2$~mag
and $A_{\rm K} (2.2\,\mu {\rm m})= 1.84 \pm 0.63$ mag.

Assuming that $A_{\rm V}/A_{\rm K} \sim 9.3$ for $R_{\rm V}=3.1$
\citep{mathis90}, we obtain that NGC\,5253~C2 is extinguished by
$17 \pm 6$ mag in the optical V-band. This value of the extinction
agrees well with the near-IR estimate by \cite{alonso04}, $18\pm5$
mag in an aperture of 0\farcs6$\times$3\arcsec. The modelling of the
near-infrared to millimetre SED by \cite{vanzi04} and \cite{alonso04}
gives comparable values for the extinction (16--21 mag). We note that
this extinction range is obtained by \cite{vanzi04} in their alternate
model which does not consider the HST optical data.
Dropping the optical flux as a constraint on the dust modelling is
more appropriate as the dominant IR emitting region (C2) and the
optically emitting regions are unrelated.
In contrast, their principal model,
which wrongly includes optical data, gives a lower extinction of
$\sim 8$~mag.

\subsubsection{Extinction in the $N$-band}
\label{sect:ext:Nband}

In order to determine the intrinsic line fluxes in the $N$-band, we
need to correct
for extinction, which, in the mid-IR, is mainly caused by silicate
absorption. Following \citet{okamoto01,okamoto03}, we have adopted a
simple model to derive the extinction in the $N$-band, where the
emission from hot dust in or at the outer edge  of the ionised region
is assumed to be extinguished by cold foreground dust along the line-of-sight.
This representation of the dust distribution is supported by the UV
study of \cite{meurer95}, who concluded that the dust geometry towards
NGC\,5253 could be described in terms of a foreground dust screen near
the starburst.
In such model, the emission of the dust at a wavelength $\lambda$
is described as:

\begin{equation}
F_\lambda = A B_\lambda(T_{\rm d}) \epsilon_\lambda {\rm
e}^{-\tau_\lambda}~,
\end{equation}

\noindent
where $A$ is a scaling factor, $T_{\rm d}$ is the temperature of the
emitting hot dust, $\epsilon_\lambda$ is the dust emissivity and
$\tau_\lambda$ is the optical depth.
We assume that the foreground dust has the same grain properties as the
emitting hot dust and hence,
$\tau_\lambda$ can be written as
$\tau_{\rm sil}\epsilon_\lambda$, where $\tau_{\rm sil}$ is the peak
optical depth of the silicate.  We have considered two extreme cases
for the optical emission of the emitting hot dust: an optically thick
approximation (where $\epsilon_\lambda = 1$) and an optically thin
approximation (where $\epsilon_\lambda \propto \tau_\lambda$).  The
adopted $\epsilon_\lambda$ is based on a mixture of carbonaceous  grains
(PAH and graphite-like) and amorphous silicate grains with $R_{\rm V} = 3.1$ developed
by \citet{weingartner01} which successfully reproduces the observed
Galactic interstellar extinction.

The 8--13 \micron\ spectrum was therefore fitted using the above
model, which has 3 free parameters: $A$, $T_{\rm d}$ and $\tau_{\rm
sil}$. We performed least-squares fits to the data.
The best fits for the optically thick and optically thin approximations
are shown in Fig.~\ref{fig:spectrum}\,b.
We obtained a
significantly better fit using the optically thick approximation
(with a resulting reduced-$\chi^2$ equal to 1.5) than using the
optically thin approximation (reduced-$\chi^2=4.8$) and therefore, we
will only consider the first case hereafter.
Figure~\ref{fig:spectrum}\,b shows the continuum-subtracted spectrum
using the optically thick model.

The best fit (optically thick approximation)
gives $A=(4.2 \pm 0.1)\times 10^{-15}$
(with $B_\lambda$ given in W\,cm$^{-2}$\,\micron$^{-1}$),
$T_{\rm d}=253 \pm 1$~K and $\tau_{\rm sil} = 0.67 \pm 0.05$.
The error analysis was done by following how $\chi^2$ changes as the parameters
in the model vary. The final errors in the best-fit parameters were
calculated by determining how much the parameters have to change to
get to the 1\% probability level.

The values of the extinction estimated at the wavelengths of the
$N$-band fine-structure
lines are given in Table~\ref{table:fluxes}, together with the line
fluxes after corrected for extinction. The extinction in magnitudes is
related to the optical depth through
$A_{\lambda}=1.086\tau_{\lambda}$.
We also give the extinction estimates and corrected line fluxes of the
mid-IR lines measured by
ISO. These line fluxes must be taken as upper limits because of the
large ISO aperture.

We can now draw the complete $2-13$~\micron\ extinction law for the
embedded cluster, where the range between 2 and $\sim$7.5~$\mu$m is
given by $A_{\lambda} = (1.84 \pm 0.63)(\lambda/2.2\,\mu {\rm
  m})^{-1.7}$ (see Sect.~\ref{sect:ext:nearIR}), and the range beyond
is given by the silicate profile by \citet{weingartner01} with $A_{\rm
  sil}=0.73 \pm 0.05$.
Since we have obtained that $A_{\rm V}=17 \pm 6$ mag (see
Sect.~\ref{sect:ext:nearIR}), we have that $A_{\rm V}/A_{\rm
  sil}=23\pm8$, in good agreement with the $A_{\rm V}/A_{\rm sil}$
ratio of $18.5\pm1.5$ found by \cite{roche84} for the local diffuse
interstellar medium.

\subsection{Ionic abundances}
\label{sect:ab}

Ionic abundances can be determined from the measured strengths of the
fine-structure and \HI\ recombination lines.
The ionic abundance of a certain ion $X^{\rm{+i}}$ with respect to hydrogen
($X^{\rm +i}/{\rm H^+}$) can be determined using the following
expression \citep[e.g.][]{rubin88}:

\begin{equation}
{X^{\rm +i} \over {\rm H^+}} = { {F_{X^{\rm +i}}/F_{{\rm HI}}} \over
{\epsilon_{X^{\rm +i}}/\epsilon_{{\rm HI}}} }~,
\end{equation}

\noindent
where $F_{X^{\rm +i}}$ and $F_{{\rm HI}}$ are the extinction-corrected
fluxes of any line produced by $X^{\rm{+i}}$ and \HI, and
$\epsilon_{X^{\rm +i}}$ and $\epsilon_{{\rm HI}}$ are their respective
emission coefficients. This expression assumes that {\em 1)} the nebula is
homogeneous with constant electron temperature and density, {\em 2)} all the
line photons emitted in the nebula escape without absorption and therefore
without causing further upward transitions and {\em 3)} the volume occupied by
$X^{\rm{+i}}$ and H$^+$ is the same.

The emission coefficients depend on
\Te, \den\ and the relevant atomic parameters \citep{martin:paperii}.
Adopting \Te=12\,000~K (see Sect.~\ref{sect:ext:nearIR}) and
\den=$5\times 10^4$~cm$^{-3}$
(see Sect.~\ref{sect:constrains}), we obtain
$\epsilon_{\rm [Ar\,III]~9.0}=8.80\times10^{-21}$,
$\epsilon_{\rm [S\,IV]~10.5}=1.41\times10^{-20}$,
$\epsilon_{\rm [Ne\,II]~12.8}=7.27\times10^{-22}$ and
$\epsilon_{\rm Br\alpha}=7.60\times10^{-27}$~erg~cm$^3$~s$^{-1}$.
Using these values for the emission coefficients and the line fluxes
listed in Table~\ref{table:fluxes}, the ionic
abundances with respect to H$^+$ we estimate for Ar$^{++}$, S$^{3+}$
and Ne$^+$ are, respectively,
$(1.0 \pm 0.4)\times10^{-6}$,
$(2.6 \pm 0.9)\times10^{-6}$ and
$(2.9 \pm 1.4)\times10^{-6}$.

Finally, we can estimate upper limits for the elemental abundances of
Ar, S and Ne using the \ArII\ 7.0, \SIII\ 18.7 and \NeIII\
15.5~\micron\ line fluxes measured
by the larger ISO aperture. The emission coefficients
of these lines for \Te=12\,000~K and \den=$5\times 10^4$~cm$^{-3}$ are
$\epsilon_{\rm [Ar\,II]~7.0}=1.22\times10^{-20}$,
$\epsilon_{\rm [S\,III]~18.7}=2.09\times10^{-21}$ and
$\epsilon_{\rm
  [Ne\,III]~15.5}=1.24\times10^{-21}$~erg~cm$^3$~s$^{-1}$. Hence,
upper limits to the ionic abundances with respect to H$^+$ of
Ar$^{+}$, S$^{++}$ and Ne$^{++}$ are, respectively,
$3.5\times10^{-7}$,
$1.6\times10^{-5}$ and
$4.7\times10^{-5}.$
Taking into account these values, we can give upper limits to the
total elemental abundances of Ar, S and Ne assuming that most of the
argon and neon is singly and doubly ionised and that most of the
sulphur appears as S$^{++}$ and S$^{3+}$. The upper limits we obtain
for Ar/H, S/H and Ne/H are, therefore,
$1.7\times10^{-6}$,
$1.9\times10^{-5}$ and
$5.1\times10^{-5}$, which are approximately
0.7, 0.9 and 0.4 times the solar elemental abundances. The upper limit
that is likely closer to the true elemental abundance is that of Ne/H
since most of the \NeIII\ line flux must come from the \HII\ region
formed by the embedded star cluster C2 (as it is the case of the \SIV\ 
line flux, cf. Sect.~\ref{sect:lines}). The upper limit obtained for Ne/H
(0.4$\times$[Ne/H]$_{\sun}$) agrees well with the global metallicity
of the galaxy ($Z\sim 0.30\,Z_{\sun}$, see
Sect.~\ref{sect:constrains}).

\section{Modelling the nebular lines}
\label{sect:modeling}

We begin with a description of the observational constraints and input
parameters for the modelling of the nebular lines.

\subsection{Observational constraints}
\label{sect:constrains}

There are several observational facts that can be used to constrain
the photoionisation models:

\paragraph{Distance}
The distance to NGC\,5253 is well known from its Cepheids. Recently,
\citet{freedman01} has derived a distance modulus of  $\mu_0=27.56 \pm
0.14$ from 17 Cepheid
candidates. This distance modulus is equivalent to a distance of $3.25
\pm 0.21$ Mpc.

\paragraph{Age}
Recently, \cite{alonso04} have estimated the age of the stellar
population of NGC\,5253~C2 based on the equivalent width (EW) of
Pa$\alpha$ ($\log$(EW(Pa$\alpha$))=$3.48 \pm 0.23$). They derive an
age of $3.3\pm1.0$~Myr. We have, however, computed the EW of
Pa$\alpha$ predicted by our photoionisation models
(Sect.~\ref{sect:input} and thereafter). The predicted EW is
practically constant during the first 6~Myr after the burst of star
formation, with a value of the order of the EW observed by
\cite{alonso04}, decreasing drastically beyond this age. Hence, based
on the observed EW, we can only obtain an upper limit for the age of
the cluster of 6~Myr. This upper limit is independent of the upper
mass cutoff of the cluster initial mass function since variations due
to different upper mass cutoffs are smaller than the quoted
observational error.  Indication of an early age comes as well from
the fact that the nuclear radio and millimetre continuum emission are
almost entirely due to thermal emission
\citep[e.g.][]{beck96,turner00}, which rules out supernova remnants or
radio supernovae as the source of the emission and place an upper
limit of about 4~Myr for the age of the cluster.  Finally, the finding
of optically thick thermal radio emission from the C2 region
\citep{turner98} interpreted as an ``ultra-dense'' \HII\ region tends
to indicate even younger ages \citep[$\la 1$~Myr;
cf.][]{kobulnicky99,vacca02} as deduced from pressure/lifetime
arguments.  However, this argument can possibly be relaxed if the
cluster mass is large enough allowing for its own gravitational
confinement, as suggested by \cite{turner03}.

\paragraph{Geometry}
Recent 7\,mm continuum observations made with the VLA including Pie
Town \citep{turner04} shows that the supernebula has a size of
$99\pm9$ mas by $39\pm4$, with a position angle of $6 \pm 4\deg$, or
1.6 pc by 0.6 pc  for the above distance of 3.25 Mpc.  The nebula is
smaller than most extra-galactic super star clusters (cf.\ Table
\ref{table:clusters}).

\paragraph{Chemical composition}
Optical and ultraviolet spectroscopy obtained with the HST
Faint Object Spectrograph at three locations in the
central \HII\ complex of NGC 5253 \citep{kobulnicky97} gives an oxygen
abundance of $12+{\rm log(O/H)}=8.16\pm0.06$ in agreement with previous
ground-based spectroscopy \citep[e.g.][]{walsh89}. Assuming a solar
oxygen abundance of $12+{\rm log(O/H)}_{\sun}=8.69$ \citep{allende01},
NGC\,5253 is a metal-poor galaxy with a metallicity $Z\sim 0.30\times
Z_{\sun}=0.006$.
This metallicity agrees well with the upper limits for the
elemental abundances of NGC\,5253~C2 derived in
Sect.~\ref{sect:ab}; in particular, it is close to the upper limit
obtained for
Ne/H, which is close to the true Ne abundance of the embedded
cluster (see also Sect.~\ref{sect:ab}).

\paragraph{Electron density} The electron density can be constrained
from radio continuum observations. Subarsecond resolution
observations of NGC\,5253~C2 indicates that the region is partially
optically thick at cm wavelengths \citep{beck96,turner00}. However, it
has been recently observed at 7~mm (43 GHz) by \cite{turner04}. The
nebular emission is optically thin at this wavelength.
The nebula contains 8.5 mJy of flux at 7~mm. Assuming that the
line-of-sight dimension is equal to the transversal size of the region
($\sim$1~pc if we take the
geometrical average of the major and minor axes of the core
as a representative size, see above), the rms electron density
(calculated using Eq. A.2.5 by
\citeauthor{panagia78} \citeyear{panagia78} ) is
\den$\sim 5\times 10^4$~cm$^{-3}$. This electron density is typical
of ultracompact \HII\ regions and it is high for an \HII\ region of
this size. Densities and sizes for Galactic \HII\ regions follow
a rather tight relation of the form
${\rm log} (n_{\rm e}) \sim 3 - 0.9 \times \log D$, where $D$ is the
diameter in pc
\citep{garay93, martin:atca:gal}. A Galactic \HII\ region of about
1~pc in diameter would have an electron density of the order of
$10^3$~cm$^{-3}$.

\paragraph{\HI\ recombination  lines}
Recombination lines of hydrogen depend on the rate of Lyman ionising
photons, $Q_{0}$. The observed and extinction-corrected
line fluxes for the Brackett $\gamma$ and $\alpha$ lines
(see Sect.~\ref{sect:ext:nearIR}) are listed in Table~\ref{table:fluxes}.

\paragraph{\HeI\ recombination lines}
\cite{lumsden94} observed the central 3\arcsec$\times$3\arcsec region
of NGC\,5253 in the K-band. They measured a Br$\gamma$ flux equal to
$0.5\times10^{-20}$~W~cm$^{-2}$. The good agreement with the
Br$\gamma$ flux measured by \cite{turner03} (see
Table~\ref{table:fluxes}) indicates that the ionised emission in the
K-band is dominated by NGC\,5253~C2. \cite{lumsden94} also
measured the 2.06~\micron\ ($2^1P-2^1S$) \HeI\ line in this band. The
observed and extinction-corrected line fluxes of this \HeI\ line are
listed in Table~\ref{table:fluxes}. The ratio \HeI\,2.06/Br$\gamma$
depends on the ratio of the He ($h\nu > 24.6$~eV) to
H ($h\nu > 13.6$~eV) ionising photons.

\paragraph{Mid-infrared fine-structure lines}

Mid-IR fine-structure lines can be used to probe the stellar SED in
the EUV range. The observed and extinction-corrected line fluxes of
the mid-infrared lines observed by TIMMI2 (\ArIII\ 9.0, \SIV\ 10.4 and
\NeII\ 12.8~\micron) are listed in Table~\ref{table:fluxes}. We also
list other mid-infrared line fluxes measured by ISO (\ArII\ 7.0,
\NeIII\ 15.5, \SIII\ 18.7 and \OIV\ 25.9~\micron),
which must be taken as upper limits because of the large ISO aperture.\\

A summary of the above observational constraints is shown in
Table~\ref{table:constrains}.

\subsection{Input parameters and assumptions}
\label{sect:input}

We compute sets of nebular models with the photoionisation code
CLOUDY\footnote{see http://thunder.pa.uky.edu/cloudy/}
\citep{ferland98} version 96.00-beta~4 using MICE\footnote{see
http://www.astro.rug.nl/$\sim$spoon/mice.html}, the IDL interface for
CLOUDY created by H. Spoon. The computation is performed for a static,
spherically symmetric, ionisation bounded gas distribution with an
inner cavity.
We assume that the gas is uniformly distributed in small clumps of
constant density over the nebular volume and occupies a fraction
$\epsilon$  of the total volume.
The shape of the stellar radiation field, the ionising
photon luminosity ($Q_0$), the electron density ($n_{\rm e}$),
the inner radius of the shell
of ionised gas ($R_{\rm in}$), the filling factor ($\epsilon$),
and the chemical composition ($Z$) are
the input parameters for the photoionisation models to predict the
intensity of the emission lines.

\begin{table}[!ht]
\caption{Observational constraints.}
  \label{table:constrains}
  \begin{center}
    \leavevmode
    \begin{tabular}[h]{lc}
      \hline \hline \\[-7pt]
    \multicolumn{1}{c}{Parameter} &
    \multicolumn{1}{c}{Value}
        \\[5pt] \hline \\[-7pt]

Br$\alpha$        & $(6.2\pm2.2)\times10^{-20}$ W cm$^{-2}$ \\
\HeI/Br$\alpha$   & $0.3\pm0.2$ \\
\ArII/Br$\alpha$  & $< 0.6$ \\
\ArIII/Br$\alpha$ & $1.2\pm0.5$ \\
\SIV/Br$\alpha$   & $4.8\pm1.7$ \\
\NeII/Br$\alpha$  & $0.3\pm0.1$ \\
\NeIII/Br$\alpha$ & $< 7.7$\\
\SIII/Br$\alpha$  & $< 4.5$\\
$R_{\rm out}$     & 0.8 pc\\
$n_{\rm e}$       & $5\times10^{4}$ cm$^{-3}$ \\
$Z$               & 0.006 \\[5pt]

\hline

    \end{tabular}
  \end{center}
\end{table}

We have used the evolutionary synthesis code
{\sc Starburst99}\footnote{see http://www.stsci.edu/science/starburst99/}
\citep{leitherer99} version 4.0 to model the integrated properties of the
stellar cluster. This code is based on stellar evolution models of the
Geneva group and uses enhanced mass loss tracks for masses above 12\,\msun\ \citep{meynet94} and standard mass loss tracks between 0.8 and 12\,\msun \citep{schaller92,schaerer93a,schaerer93b,charbonnel93}. {\sc Starburst99} follows the evolution in the H-R
diagram of a stellar population whose composition is specified by a
stellar initial mass function (IMF). At a given age, the integrated
SED is obtained by summing over the contributions of all stars present
and is built using {\em 1)} the non-LTE {\sc WM-Basic} stellar models
\citep{pauldrach01, smith02} for O stars, which take into account the effects
of stellar winds and line blanketing, {\em 2)} the fully line-blanketed models
calculated with the CMFGEN code \citep{hillier98} for Wolf-Rayet (WR) stars
and {\em 3)} the plane-parallel
LTE models by \cite{kurucz93} for the remaining stars that contribute to
the continuum.
We assume an instantaneous burst of star formation, a
\cite{salpeter55} IMF with exponent $\alpha=2.35$ ($dN/d\ln m \propto
m^{1-\alpha}$), a lower mass cutoff
$M_{\rm low}=1$~M$_{\sun}$ and an upper mass cutoff $M_{\rm up}$ set to 30,
50 and 100~M$_{\sun}$. We also assume that the stars evolve from the main
sequence following the $Z=0.004$ and $Z=0.008$ stellar tracks (stellar
tracks with $Z=0.006$ are not currently available).
We present models every 0.5 Myr for 10 Myr after the burst of star
formation.

The chemical composition of the gas is set to $Z=0.30\times
Z_{\sun}=0.006$ (see Sect.~\ref{sect:constrains}).
We have adjusted the helium abundance $Y$ according to
$Y=Y_{\rm p}+(\Delta Y/\Delta Z)Z$, where $Y_{\rm p}=0.24$ is the
primordial helium abundance \citep{audouze87} and
$(\Delta Y/\Delta Z)=3$ is an observed constant \citep{pagel92}. To
arrive at the appropriate metal abundances, we have simply scaled the solar
values.

A change of the rate of Lyman ionising photons,
electron density, inner radius of the shell and/or
filling factor is equivalent to a change of the
ionisation parameter ($U$) defined by CLOUDY as:

\begin{equation}
U=Q_{0}/(4\pi R_{\rm in}^2 n_{\rm H}c)~,
\label{eq:u}
\end{equation}

\noindent
where $R_{\rm in}$ is the distance between the ionising cluster and
the illuminated
surface of the gas shell and $n_{\rm H}$ is the total hydrogen density
(ionised, neutral and molecular). Basically, $U$ is the
dimensionless ratio of hydrogen ionising photons to total hydrogen
density.
We fix an electron density
equal to the rms electron density derived from radio continuum
observations ($5\times 10^4$~cm$^{-3}$, see Sect.~\ref{sect:constrains}).
$Q_0$ is fixed individually for every photoionisation model
so that the Br$\alpha$ line flux is reproduced within 5\%. Test models
give that the resulting $Q_0$ varies slightly with the cluster age,
with values between $1.6\times10^{52}$ and $2\times10^{52}$~s$^{-1}$, in
fair agreement with the $\sim1\times10^{52}$~s$^{-1}$ estimated from
the 7\,mm observations presented by \cite{turner04}.

Once we have fixed $n_{\rm e}$ and $Q_0$, there is still a degeneracy between
the inner and outer radius of the \HII\ region and its filling factor.
From \cite{osterbrock89}, we have that:

\begin{equation}
Q_0 = {{4 \pi} \over 3} (R_{\rm out}^3 - R_{\rm in}^3) \epsilon
n_{\rm e}^2 \alpha_{\rm B}~,
\label{eq:q}
\end{equation}

\noindent
where the rate of ionising photons emitted by the cluster
($Q_0$) just balances the total number of \HI\ recombinations to
excited levels within the ionised volume
$4\pi(R_{\rm out}^3 - R_{\rm in}^3)/3$, where $R_{\rm out}$ is the
outer radius of the gas shell. The parameter
$\alpha_{\rm B}$ is the
case B \HI\ recombination coefficient to all levels $\ge 2$.
Using Eq.~\ref{eq:q}, we can express $U$ (Eq.~\ref{eq:u}) as
a function of $R_{\rm in}$, $R_{\rm out}$ and $\epsilon$ as follows:

\begin{equation}
U= { \alpha_{\rm B} \over {3c} } n_{\rm e} \epsilon
\left ( { {R_{\rm out}^3 - R_{\rm in}^3} \over {R_{\rm in}^2} } \right )~,
\label{eq:all}
\end{equation}

\noindent
where we have assumed that $n_{\rm H} \simeq n_{\rm e}$.
The size of the source
observed by \cite{turner04} is $1.6 \times 0.6$~pc
(see Sect.~\ref{sect:constrains}). Hence, $R_{\rm out}$ cannot be larger
than 0.8~pc. The outer radius of
the model comes defined by the position where the ratio of
electron to total hydrogen densities reaches a value of 0.90, i.e.
near the H$^+$--H$^0$ ionisation front.
$R_{\rm in}$ and $\epsilon$, however,
cannot be constrained from observations, but we can set them to reasonable
values.

Equation~\ref{eq:q} can be re-written as:

\begin{equation}
(1 - x^3) R_{\rm out}^3 \epsilon =
{ {3 Q_0} \over {4 \pi n_{\rm e}^2 \alpha_{\rm B}} } =
{ {2.14\times10^{64}} \over {n_{\rm e}^2} }  = 
{ {\beta} \over {n_{\rm e}^2} }~,
\label{eq:x}
\end{equation}

\noindent
where $R_{\rm out}$ is in pc and $R_{\rm in} = xR_{\rm out}$, with
$0 < x < 1$. We have adopted
$Q_0=2\times10^{52}$~s$^{-1}$. Physical solutions of this equation are
those that satisfy the condition $\epsilon \le 1$, which is equivalent to

\begin{equation}
x \le 
\left ( 1 - { {\beta} \over {R_{\rm out}^3 n_{\rm e}^2} } \right )^{1/3}~.
\label{eq:condition}
\end{equation}

\noindent
This condition leads to $x \le 0.75$ for \den=$5\times10^4$~cm$^{-3}$
and $R_{\rm out}=0.8$~pc. We note that this condition does not
allow values of $R_{\rm out} < 0.7$ pc and \den $< 3.7\times10^4$~cm$^{-3}$.
The minimum value allowed for the filling factor is
$\epsilon (x \rightarrow 0) = 0.6$.  Hence, the final constraints on
the geometry are $R_{\rm out}=0.8$~pc, $R_{\rm in} \leq 0.6$~pc and
$0.6 \leq \epsilon \leq 1$. We remind that these three parameters are
not independent, but have to satisfy the condition expressed by
Eq.~\ref{eq:x}.

   \begin{figure}[!ht]
   \centering \includegraphics[width=8.8cm]{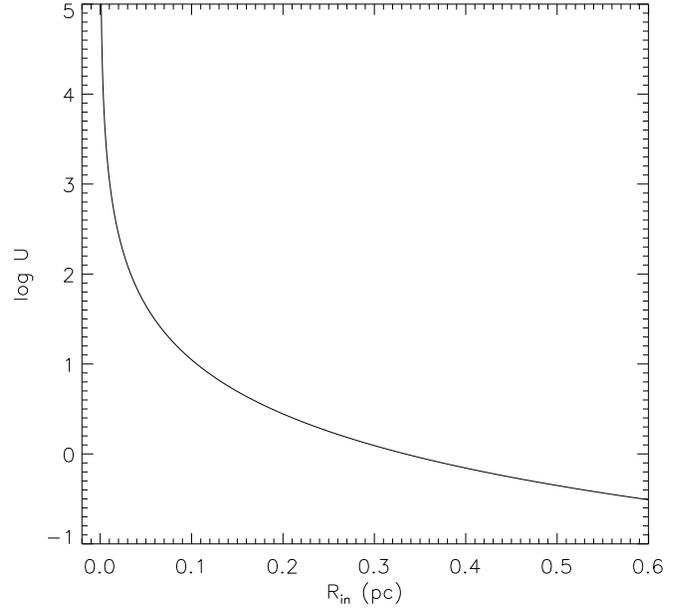}
   \caption{Variation of ionisation parameter $U$ with inner radius
   ($R_{\rm in}$)
   for an \HII\ region with
   $Q_0=2\times10^{52}$~s$^{-1}$ and \den=$5\times10^4$~cm$^{-3}$.
   $R_{\rm in}=0.6$~pc is the maximum value allowed for the inner radius of
   an \HII\ region with $R_{\rm out}=0.8$~pc.}
        \label{fig:test_u}
   \end{figure}

\begin{table}[!ht]
\caption{Input parameters of the fiducial model.}
  \label{table:input}
  \begin{center}
    \leavevmode
    \begin{tabular}[h]{ll}
      \hline \hline \\[-7pt]

      \multicolumn{1}{c}{Parameter} &
      \multicolumn{1}{c}{Value} \\[5pt] \hline \\[-7pt]

      $R_{\rm in}$  & 0.6~pc                    \\
      $R_{\rm out}$ & 0.8~pc                    \\
      $\epsilon$    & 1                         \\
      \den          & $5\times10^4$~cm$^{-3}$   \\
      $Z_{\rm gas}$ & 0.3$Z_{\sun}$             \\[5pt]

\hline

    \end{tabular}
  \end{center}
\end{table}

We evaluate now how the above variations of $R_{\rm in}$ and $\epsilon$
modify the ionisation parameter $U$. Fig.~\ref{fig:test_u} shows the
variation of $U$ for an \HII\ region with
$Q_0=2\times10^{52}$~s$^{-1}$ and \den=$5\times10^4$~cm$^{-3}$
(Eq.~\ref{eq:u}). $R_{\rm in}=0.6$~pc (which corresponds to a filling
factor $\epsilon=1$) is the maximum value allowed for the inner radius
if $R_{\rm out}=0.8$~pc.
The ionisation parameter for a nebula with $R_{\rm in}=0.6$~pc and
$\epsilon = 1$ is, according to the figure, log$U \simeq -0.5$~dex.
However, regions with smaller inner radii will have larger ionisation
parameters.

We will adopt a reference geometry with $R_{\rm in}=0.6$~pc and
$\epsilon = 1$ for the main bulk of models presented hereafter. A
summary of the input parameters of these fiducial models is presented
in Table~\ref{table:input}. The effect of varying $U$
(i.e. $R_{\rm in}$ and $\epsilon$) and other parameters such as the slope 
of the
IMF, the nebular metal content, the internal dust content and
density structure
will be discussed in the following section.

\subsection{Results from the photoionisation models}
\label{sect:models}

\begin{figure*}[!ht]
   \centering \includegraphics[width=14.5cm]{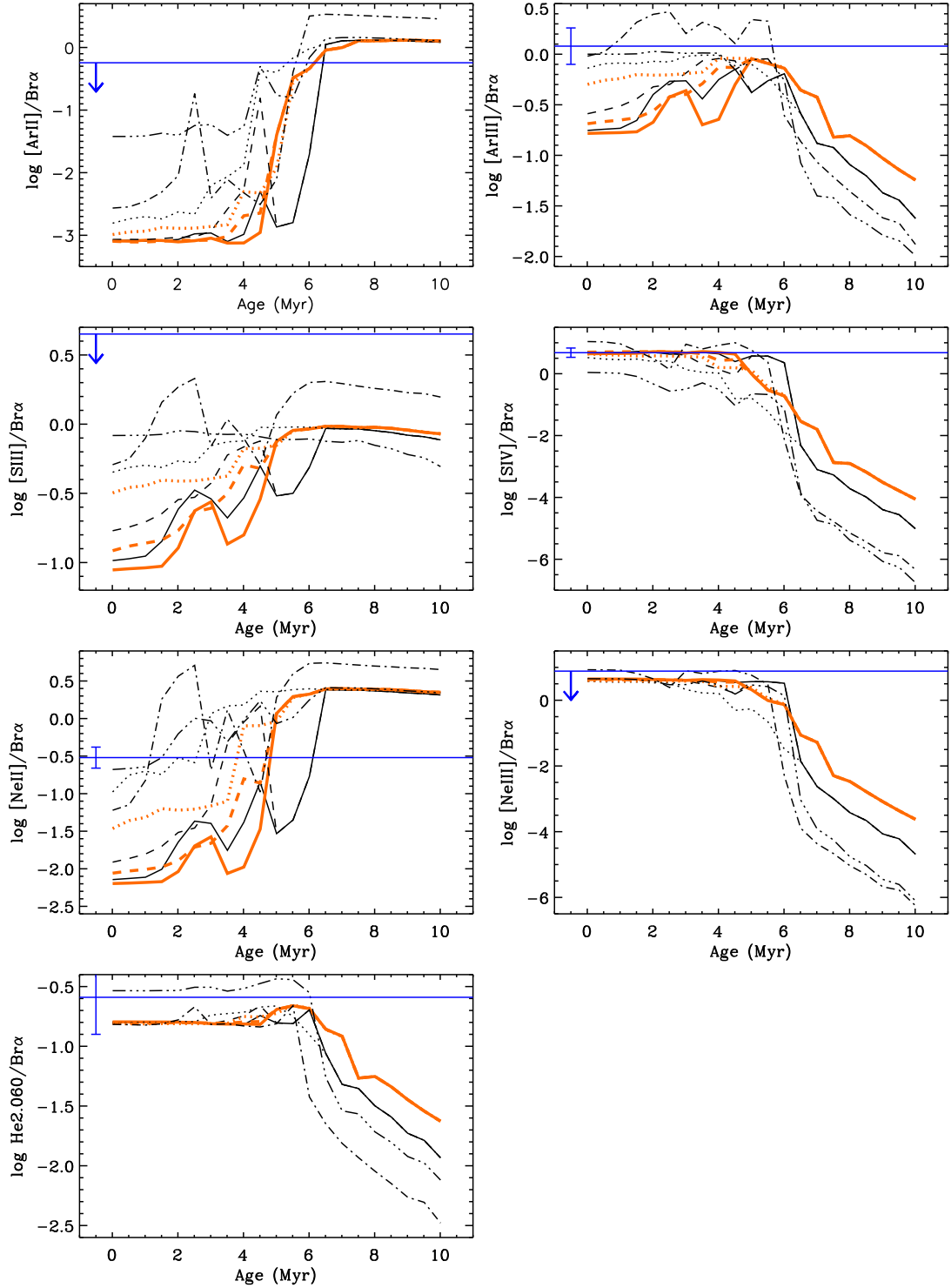}
   \caption{Variation of the selected emission line ratios
        as a function of starburst age.
     The models reproduce the observed Br$\alpha$ line flux within 5\%.
     The observed ratio is indicated by an horizontal line and its
     uncertainty by an error
     bar on the left side.
     Upper limits correspond to a mix of ISO and small aperture
     measurements (see Table~\ref{table:fluxes}).
     The legend of the figures is the following:
     lines in black correspond to models with Z=0.008;
     lines in light colour to models with Z=0.004;
     solid lines correspond to models with $M_{\rm up}=100M_{\sun}$;
     dashed lines to models with $M_{\rm up}=50M_{\sun}$ and
     dotted lines to models with $M_{\rm up}=30M_{\sun}$.
     Dashed-dotted lines correspond to models with
     solar metallicity and $M_{\rm up}=100M_{\sun}$.
     The nebular parameters of all these models are $R_{\rm in}=0.6$~pc,
     $R_{\rm out}=0.8$~pc,
     $\epsilon=1$ and \den$=5\times10^4$~cm$^{-3}$.
     Dashed-dotted-dotted lines correspond to models with $M_{\rm
     up}=100M_{\sun}$,
     $Z=0.008$ and the following nebular parameters: $R_{\rm in} \sim
     R_{\rm out}=4.5$~pc,
     $\epsilon=0.1$ and \den$=5\times10^4$~cm$^{-3}$.
     }
         \label{fig:cloudy_1}
   \end{figure*}

\begin{figure*}[!ht]
   \centering \includegraphics[width=18cm]{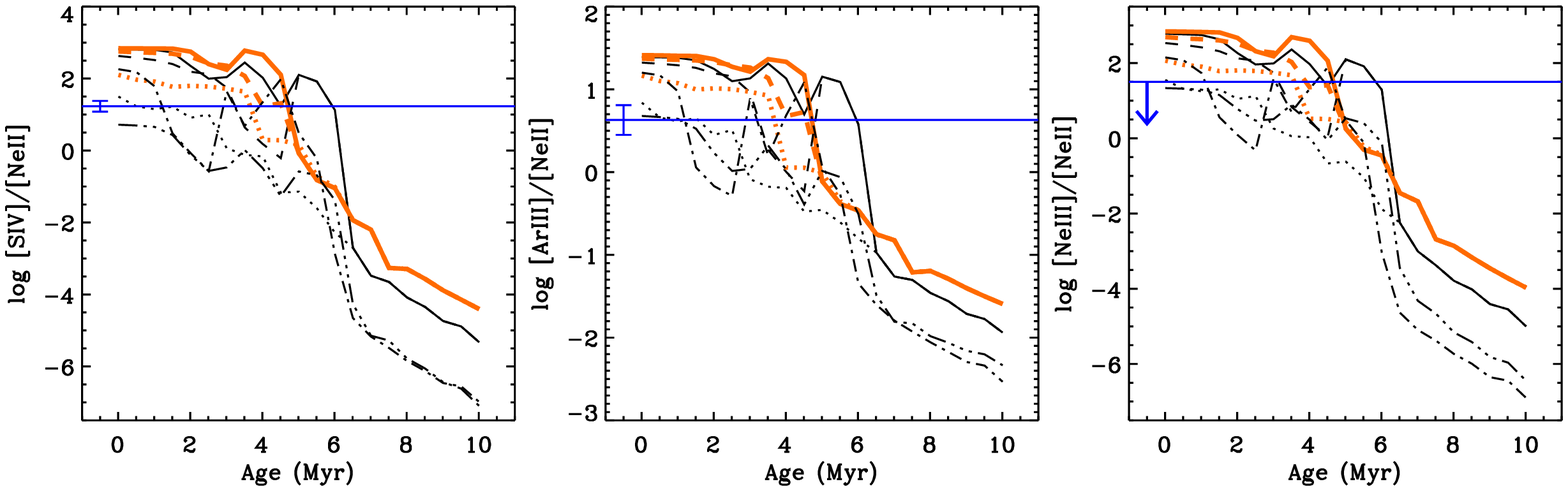}
   \caption{Variation of the mid-infrared line ratios as a
     function of starburst age. The observed value is indicated
     by an horizontal line and its uncertainty by an error bar on the
     left side. Upper limits correspond to a mix of ISO and small aperture
     measurements (see Table~\ref{table:fluxes}).
     Models are plotted using the same code as in Fig.~\ref{fig:cloudy_1}.
}
        \label{fig:cloudy_2}
   \end{figure*}

Figures~\ref{fig:cloudy_1} and \ref{fig:cloudy_2} show the main results of
the photoionisation models compared to the (extinction-corrected)
observed values listed in Table~\ref{table:fluxes}.
Figure~\ref{fig:cloudy_1} shows the variation of the seven independent 
constraints from
Table~\ref{table:fluxes}
(\ArII/Br$\alpha$,
\ArIII/Br$\alpha$,
\SIII/Br$\alpha$,
\SIV/Br$\alpha$,
\NeII/Br$\alpha$,
\NeIII/Br$\alpha$ and
\HeI\,2.06/Br$\alpha$)
as a function of time since the burst of star formation.
We note that these models reproduce the observed Br$\alpha$ line flux
within 5\%.
Figure~\ref{fig:cloudy_2} shows the variation of the mid-infrared line
ratios \SIV/\NeII, \ArIII/\NeII\ and \NeIII/\NeII.

In agreement with earlier combined starburst and photoionisation models
\citep[e.g.][]{schaerer99,schaerer99a,rigby04},
we observe
that ratios such as \SIV/\NeII, \ArIII/\NeII\ and \NeIII/\NeII\
(see Fig.~\ref{fig:cloudy_2})
decrease smoothly during approximately the first 3 Myr as
the hottest O stars leave the main sequence, increasing again from 3 to
4--6 Myr as a consequence of the contribution of the WR stars and
the remaining main sequence stars to finally drop afterwards.
This clearly reflects the importance of WR stars, which contribute
to the ionisation of the nebula after the hottest O stars have left the
main sequence. This behaviour is especially noticeable in the case of models
with $M_{\rm up}=100~M_{\sun}$.
This contribution of the WR stars is also observable
in Fig.~\ref{fig:cloudy_1}.

These figures clearly show the dependence
of these line ratios on the stellar content and metallicity.
A higher $M_{\rm up}$ implies a harder stellar spectrum and consequently,
a higher degree of excitation. Similarly, a lower stellar metallicity
leads as well to a harder stellar spectrum.

We will focus first on the subsolar predictions.  The comparison
between the results of these photoionisation models and the
(extinction-corrected) observed line ratios clearly shows that it is
not possible to constrain at the same time both the age and $M_{\rm
up}$ of the stellar cluster. However, some conclusions can be derived.
First, it is shown that NGC\,5253~C2 has an age of less than 6
Myr. All model predictions beyond this age fall far from the
observations. A cluster with $M_{\rm up}=100\,M_{\sun}$ reproduces
well all the observed lines for an age of $\sim 5-6$~Myr.  Second, a
closer inspection reveals that younger clusters fit the observed line
fluxes only when the upper mass cutoff is smaller. For instance, a
cluster younger than about 4 Myr can only reproduce the observed line
ratios if its $M_{\rm up}$ is lower than 50$M_{\sun}$.  An even
younger cluster of about 1 Myr could only fit the observations if its
$M_{\rm up}$ is lower than 30$M_{\sun}$.

The large spectral range covered by ISO included line fluxes
such as \ArII\ 7.0, \NeIII\ 15.5 and \SIII\ 18.7~\micron. These lines
fluxes (listed in Table~\ref{table:fluxes}) are upper limits to the
true fluxes emitted by the nebula because of the large ISO
aperture. They are compared to the model predictions in
Fig.~\ref{fig:cloudy_1}.  As it was already pointed out in
Sect.~\ref{sect:lines}, the comparison of the models with the observed
line fluxes reveal that most of the emission from the high-excitation
\NeIII\ line (the ionisation potential of Ne$^{++}$ is 41~eV) comes
from C2. However, other sources included in the ISO
aperture must contribute to the low-excitation \ArII\ and \SIII\ line
fluxes (Ar$^+$ and S$^{++}$ have ionisation potentials of 15.8 and
23.3~eV) since the models predict much lower fluxes. Indeed, these
fluxes must be taken as upper limits. Considering that practically all
the \NeIII\ line flux measured by ISO is emitted by the embedded SSC C2,
we can confirm that the upper limit to the elemental abundance
of neon estimated in Sect.~\ref{sect:ab} ($0.4\times$[Ne/H]$_{\sun}$)
is close to the true neon abundance of the \HII\ region. Since neon is
a primary product of the stellar synthesis of elements and its
abundance closely follows that of oxygen, we can affirm that the \HII\
region excited by NGC\,5253~C2 is metal-poor with a metallicity close
to that measured for the galaxy.

The effect of varying the ionisation parameter $U$
(i.e. $R_{\rm in}$ and/or $\epsilon$) and other parameters such as the
slope of the IMF, the nebular and stellar metal content, the internal dust
content and the density structure, as well as the effect of a matter
bounded geometry are discussed below.

\subsubsection{The ionisation factor}
\label{sect:u}

We have adopted $R_{\rm out}=0.8$~pc, $R_{\rm in}=0.6$~pc and
$\epsilon=1$ for the models presented here. As mentioned in
Sect.~\ref{sect:input}, possible solutions for the geometry of the
nebula include lower values of $R_{\rm in}$ and $\epsilon$. Such
solutions will produce an increase of the ionisation factor, as was
shown in Fig.~\ref{fig:test_u}, and of the general ionisation level of
the nebula. Hence, the models presented in Figs.~\ref{fig:cloudy_1} and
\ref{fig:cloudy_2} provide an upper limit to the upper mass cutoff for
a given starburst age in the case of the nebula having a smaller inner
radius (or a larger ionisation parameter).

Another aspect that can be analysed with respect to the ionisation
factor is that of how it can be lower. A decrease of $Q_0$ or an
increase of the size of the \HII\ region will indeed lower the
parameter $U$. Assuming that $Q_0$ is well constrained from the
observed Br$\alpha$ line flux, other geometries must satisfy the
condition given by Eq.~\ref{eq:condition}. An upper limit to the size
of the nebula is given by the 11.7~\micron\ image of \cite{gorjian01},
where the source is seeing as an unresolved object with a diameter of
0\farcs58 ($\sim 9$~pc). We will take then $R_{\rm out}=4.5$~pc as the
largest possible outer radius of the nebula. Using this outer radius,
the condition given by Eq.~\ref{eq:condition} requires an electron
density larger than 2830~cm$^{-3}$. The lowest value for the
ionisation parameter is obtained considering the largest possible
values for the inner radius and electron density (see
Eq.~\ref{eq:u}). The electron density cannot be larger than that
derived from the 7~mm observations ($5\times10^4$~cm$^{-3}$) for a
much smaller and compact \HII\ region. A nebula with $R_{\rm in} \sim
R_{\rm out}=4.5$~pc and \den=$5\times10^4$~cm$^{-3}$ will have an
ionisation parameter equal to $\log U \approx -2.2$. The filling
factor of a nebular with this geometry is constrained by
Eq.~\ref{eq:condition}: $\epsilon (R_{\rm in} \rightarrow R_{\rm out})
\simeq 0.1$.

We have run a set of photoionisation models using this geometry
(\den=$5\times10^4$~cm$^{-3}$, $R_{\rm in}=4.5$~pc and $\epsilon=0.1$)
and stellar SEDs with $M_{\rm up}=100M_{\sun}$ and $Z=0.008$. These
models are plotted by dashed-dotted-dotted lines in
Figs.~\ref{fig:cloudy_1} and \ref{fig:cloudy_2}.
They reproduce well the observed \HeI\ line flux
and the \NeIII\ measured by ISO. Regarding the IR lines observed by
TIMMI2, the \NeII\ and \ArIII\ line fluxes are well reproduced by star
clusters younger than 4 Myr, while the \SIV\ line flux is
underpredicted by a factor of about 4--15 in this age range.

Stellar clusters with an upper mass cutoff of $100M_{\sun}$ can
marginally reproduce all the observed line fluxes when the radius of
the \HII\ region is allowed to be as large as 4.5~pc.
However, the Lyman continuum photon rate we estimate from the
Br$\alpha$ line flux ($\sim 2\times10^{52}$~s$^{-1}$) is of the order
of the value obtained from the 7~mm radio observations ($\sim
1\times10^{52}$~s$^{-1}$ when a distance of 3.25~Mpc is assumed) we
use to constrain the size of the nebula. Hence, it does not seem
plausible that the nebula has a radius much larger than the value of
0.8~pc we have considered.

\subsubsection{The power law index of the IMF}
\label{sect:alpha}

We have adopted a Salpeter IMF with a power law index
$\alpha=2.35$. However, for massive stars, \cite{scalo86} finds a
steeper IMF with $\alpha=2.7$. Correcting for
binaries can also lead
to systematically larger exponents \citep[$\alpha \simeq 2.4-3$;][]{kroupa02}.
A larger value ($\alpha \simeq 3.0 \pm 0.1$) is also
suggested by a completely independent but indirect approach based on
the FIR luminosity function of IRAS point sources with colours of
ultracompact \HII\ regions and CS(2--1) detections in the Galactic
disk \citep{casassus00}.

We analyse here the effect of a steeper IMF on the photoionisation
model results. In order to do so, we have run a set of photoionisation
models with the reference parameters listed in Table~\ref{table:input}
and stellar SEDs with $M_{\rm up}=100\,M_{\sun}$, $Z=0.008$ and
$\alpha=3.0$. The effect of a steeper IMF with $\alpha=3.0$ on the
considered line ratios is not very significant. In the age interval
0--6 Myr, the predicted \ArIII\ and \NeII\ line fluxes increase by
factors $<1.2$ and $<2.1$, respectively, while the \SIV\ line
flux decreases by a factor $>0.8$. Hence, there is a softening of the
cluster SED because of the smaller number of very massive stars;
however, the effect on the mid-IR lines is small.
Concluding, an IMF with a steeper slope does not reconcile
stellar clusters with $M_{\rm up}=100\,M_{\sun}$ and age $<$ 4~Myr
with the observed line fluxes.

The effect of a top-heavy IMF (i.e. with a smaller power law index)
would be a hardening of the SED because of the increased number of
very massive stars. Hence, the fiducial models presented in
Figs.~\ref{fig:cloudy_1} and \ref{fig:cloudy_2} (with a Salpeter index,
$\alpha=2.35$) provide an upper limit to the upper mass cutoff for a
given starburst age in the case of an IMF with $\alpha < 2.35$.
There are only minor indications of clusters with top-heavy IMF's. For
instance, \cite{figer99} find an index $\alpha \sim 1.6$ for the
massive Arches and Quintuplet clusters situated near the Galactic
centre. There also indications of a top-heavy IMF in the massive M82-F
cluster located in the starburst galaxy M82. The mass-to-light ratio
derived towards this cluster \citep{smith01} is significantly smaller
that the ratio expected from a normal IMF, indicating a deficit of
low-mass stars and a top-heavy IMF. However, other well-studied and
resolved massive clusters such as 30\,Doradus \citep{selman99} and
NGC\,3603 \citep{eisenhauer98} have an IMF power law index close to
the Salpeter value.

\subsubsection{Metal enrichment}
\label{sect:solar}

The effect of a possible metal enrichment of the parental molecular cloud
is discussed here. We have considered an \HII\ region with the same
geometrical parameters as in the fiducial models (see
Table~\ref{table:input}) but considering
solar stellar (and nebular) metallicity. The predictions of the
photoionisation models are plotted in Figs.~\ref{fig:cloudy_1} and
\ref{fig:cloudy_2} by a dashed-dotted line.
The general effect of increasing the metallicity is
softening the SED and lowering the general ionisation level of the
nebular gas. A young stellar cluster with $M_{\rm up}=100 M_{\sun}$ and
solar metallicity can reproduce the observed line fluxes for ages $< 4$~Myr.

However, the upper limits for the elemental abundances estimated in
Sect.~\ref{sect:ab} (especially that for Ne/H, which must be close to
the true neon abundance of the \HII\ region, see the discussion above)
do not give evidences of a possible metal enrichment. This test is
anyway useful to show how wrong assumptions on the metallicity can
lead to wrong conclusions on the cluster properties.

\subsubsection{Internal dust}
\label{sect:dust}

The effect of adding internal dust in the nebula is to increase the
absorption of ionising photons and to modify the shape of the ionising
spectrum.
The new CLOUDY version 96.00--beta~5 presents a very detailed treatment of the
grain physics, where the heating and cooling of the gas by grain
photoionisation-recombination
is determined consistently and the grain size distribution is
resolved. Resolving the size
distribution can lead to significant changes in the emitted spectrum
at IR wavelengths.
CLOUDY allows the use of different grain size distribution
functions. We will consider
here two of these distributions: the ISM type, which reproduces the
observed ratio of total to selective extinction (i.e. $R_{\rm
  V}=3.1$), and the Orion size distribution, which is deficient in
small particles and reproduces the larger $R_{\rm V}$ observed in Orion.

We have run several photoionisation models using a stellar SED with
$M_{\rm up}=100M_{\sun}$,
1 Myr and $Z=0.008$, and the nebular parameters of the fiducial models
(see Table~\ref{table:input}).
We have considered internal dust following the ISM and Orion size
distributions with an abundance 0.1 times their standard abundance.
This is basically a scale factor used to multiply the stored grain
opacities. For both types of dust, we have fixed $Q_0$ such that we
reproduce the Br$\alpha$ line flux within 5\% (we note that the flux
of Br$\alpha$ have only been corrected for foreground extinction).

\begin{table*}[!ht]
\caption{Comparison between predictions of models with different density profiles
(see Sect.~\ref{sect:dprofile}) and the observations. These models have been run
using a stellar SED with \mup=100\,\msun, 1~Myr and $Z=0.008$. They reproduce the observed Br$\alpha$ line flux within 5\%.}
  \label{table:den}
  \begin{center}
    \leavevmode
    \begin{tabular}[h]{lccrrrrrr}
      \hline \hline \\[-7pt]

      \multicolumn{1}{c}{Model} &
      \multicolumn{1}{c}{$n_{\rm o}$} &
      \multicolumn{1}{c}{$\gamma$} &
      \multicolumn{1}{c}{log \ArIII/Br$\gamma$} &
      \multicolumn{1}{c}{log \SIV/Br$\gamma$} &
      \multicolumn{1}{c}{log \NeII/Br$\gamma$} &
      \multicolumn{1}{c}{log \SIV/\NeII} &
      \multicolumn{1}{c}{$R_{\rm in}$ (pc)} &
      \multicolumn{1}{c}{$\Delta L^\star$ (pc)}
      \\[5pt] \hline \\[-7pt]

\#1 & $5\times10^4$ & 0 & $-0.73$ &   0.68  & $-2.11$ &   2.80 & 0.6 & 0.19 \\
\#2 & $5\times10^4$ & 1 & $-0.76$ &   0.75  & $-2.17$ &   2.92 & 0.6 & 0.26 \\
\#3 & $5\times10^4$ & 2 & $-0.80$ &   0.84  & $-2.23$ &   3.08 & 0.6 & 0.41 \\[5pt]

\#4 & $5\times10^3$ & 0 & $-0.59$ &   1.14  & $-2.00$ &   3.15 & 0.45& 2.48 \\
\#5 & $5\times10^3$ & 1 & $-0.40$ &   1.24  & $-1.72$ &   2.95 & 0.45& 41.3 \\[5pt]

\#6 & $5\times10^5$ & 0 & $-0.52$ & $-0.38$ & $-1.35$ &   0.97 & 0.8 &$1.6\times10^{-3}$ \\
\#7 & $10^7$        & 0 & $-1.50$ & $-2.86$ & $-1.25$ & $-1.61$& 0.8 &$4.8\times10^{-6}$\\[5pt]

Obs & & & $0.08\pm0.18$ & $0.68\pm0.15$ & $-0.52\pm0.14$ & $1.23\pm0.15$\\[5pt]

\hline

      \end{tabular}
  \end{center}

$^\star$ Thickness of the shell.

\end{table*}

A first result of these dusty models is that, in order to reproduce
the observed Br$\alpha$ flux, we get a larger $Q_0$
with respect to the value obtained for the non-dusty models. In
particular, $Q_0$ increases to about $4.0\times10^{53}$~s$^{-1}$ and
$6.6\times10^{52}$~s$^{-1}$ for the models with ISM and Orion-like
dust grains, respectively. With respect to the mid-IR line fluxes, the
models give the following ratios: log(\NeII/Br$\alpha$)=$-2.7$ and
$-2.3$, log(\ArIII/Br$\alpha$)=$-1.3$ and $-1.0$,
log(\SIV/Br$\alpha$)= 0.5 and 0.7, log(\SIV/\NeII)=3.2 and 3.0, and
log(\ArIII/\NeII)=1.4 and 1.3. Hence, we see that the general
ionisation level increases with respect to the non-dusty models,
increasing even more the difference with the observed line fluxes.
This is because the absorption cross section per H nucleon decreases
for energies above 13.6~eV and consequently, the presence of dust favours the
ionisation of He and other ions with respect to H.

As a conclusion, the presence of internal dust does not reconcile the
stellar clusters with $M_{\rm up}=100M_{\sun}$ and age $< 4$~Myr with the 
observed line
fluxes. The non-dusty models presented in Figs.~\ref{fig:cloudy_1} and
\ref{fig:cloudy_2} provide an upper limit to the upper mass cutoff for
a given starburst age in the case of the presence of internal dust.

\subsubsection{Density profile}
\label{sect:dprofile}

We have considered so far photoionisation models with a constant
electron density equal to $5\times10^4$~cm$^{-3}$.  We will analyse
now the effect of an electron density profile which is allowed to vary
as a power law of the form $n_{\rm e}=n_{\rm o} \times (R/R_{\rm
in})^{-\gamma}$, where $n_{\rm o}$ is the electron density at the
illuminated face of the nebula and $R$ is the nebular radius.  We have
run a set of models with different values of $n_{\rm o}$ and $\gamma$
using a stellar SED with \mup=100\,\msun, 1~Myr and $Z=0.008$.
Predictions of these models are shown in Table~\ref{table:den}, where
they are compared with the fiducial model (\#1) and the observations.

The first test, which comprises models \#2 and \#3, analyses the
effect of a density gradient with an inner density equal to
$5\times10^4$~cm$^{-3}$. A density gradient produces a slight increase
of the \SIV/\NeII\ line ratio and a thickening of the ionised shell
when compared with the fiducial model \#1. The discrepancy with the
observations is even higher. This increase in \SIV/\NeII\ is
principally due to the fact that the \SIV\ 10.5~\micron\ line has a
lower critical density ($3.7\times10^4$~cm$^{-3}$) than the \NeII\
12.8~\micron\ line ($6.1\times10^5$~cm$^{-3}$).  Hence, a density
gradient favours the enhancement of the \SIV\ line with respect to the
\NeII\ line (and also with respect to the \ArIII\ line, with a
critical density equal to $2.1\times10^5$~cm$^{-3}$).

The second test (models \#4 and \#5) studies the effect of decreasing
the inner density, $n_{\rm o}$. As noted in Sect.~\ref{sect:input}, an
\HII\ region with $Q_0=2\times10^{52}$~s$^{-1}$ and $R_{\rm
out}=0.8$~pc does not allow values of the electron density lower than
$3.7\times10^4$~cm$^{-3}$. As it is shown by Eq.~\ref{eq:condition},
the smallest \HII\ region with $Q_0=2\times10^{52}$~s$^{-1}$ and
$n_{\rm e}=5\times10^3$~cm$^{-3}$ would have an outer radius of $\sim
3$~pc (with $R_{\rm in}=0.45$~pc), about 4 times larger than what is
observed at radio wavelengths. In terms of the line fluxes, these
models predict higher fluxes and a general increase of the excitation
of the gas (as measured by \SIV/\NeII). The discrepancy with the
observations is, again, higher than that of the fiducial model.

The third and final test (models \#6 and \#7) analyses the effect of
increasing the inner density. We study two cases: $n_{\rm o} =
5\times10^5$ and $10^7$~cm$^{-3}$. We only consider constant
densities. At these high densities, the thickness of the shell is very
small and density gradients will not produce any effect on the
predictions. These models predict an important decrease of the
\SIV/\NeII\ ratio. This effect goes in the right direction to
reconcile the models with \mup=100\,\msun\ and the observations. The
model with $n_{\rm o} = 5\times10^5$~cm$^{-3}$ even reproduces rather
well the \SIV/\NeII\ line ratio; however, it fails to reproduce the
individual fluxes of \ArIII, \SIV\ and \NeII.  The increase of the
density up to $10^7$~cm$^{-3}$ comes along with an important weakening
of the \SIV\ line flux. This weakening is also seen in the \ArIII\
line. On the contrary, the flux of the \NeII\ slightly increases. This
is because, at this high density, collisional de-excitation of the
upper levels of the \ArIII\ and \SIV\ lines is important, while the
\NeII\ line, with a higher critical density, is much less
affected. Hence, an increase of the density, even though leads to a
lower ionisation level of the nebula, fails to reproduce the observed
\ArIII\ and \SIV\ line fluxes, while the \NeII\ line flux, although
higher, is still underpredicted.

In conclusion, variations in the density and/or density profile do not
reconcile the models with \mup=100\,\msun\ and 1~Myr with the
observations. The models presented in Figs.~\ref{fig:cloudy_1} and
\ref{fig:cloudy_2} provide an upper limit to the upper mass cutoff for
a given starburst age in the case of power-law density profiles with
inner densities lower than that assumed in the fiducial models
($5\times10^4$~cm$^{-3}$). We have also shown that even though an
increase of the density leads to lower values of the \SIV/\NeII\
ratio, such models largely underpredict the \ArIII, \SIV\ and \NeII\
line fluxes.

\subsubsection{Matter bounded geometry}
\label{sect:matter}

The nebular models presented here have an ionisation (or radiation)
bounded geometry, where the outer limit of the \HII\ region is defined
by a hydrogen front. In this geometry, the \HII\ region is optically
thick to the hydrogen ionising continuum and has absorbed nearly all
of it. \HII\ regions can be also matter (or density) bounded if the
outer limit is set by the edge of the cloud. In this case, the \HII\
region is optically thin to the incident continuum.
The effect of a matter bounded geometry will be then to cut off the
external parts of the nebula, i.e., the lower excitation regions, producing a
nebula with a higher ionisation level.
Consequently, if the nebular were matter-bounded, the upper mass cutoff of the IMF
would need to be revised downward in comparison to the models
discussed above (Figs.~\ref{fig:cloudy_1} and \ref{fig:cloudy_2}).

\subsection{On the \OIV\ emission in NGC 5253}
\label{sect:oiv}

The high-excitation \OIV\ line at 25.9~\micron\ has been observed in
active galactic nuclei and in some starburst galaxies
including NGC 5253 \citep[e.g.][]{genzel98,lutz98}.
The ionisation edge for the creation of \ion{O}{iv} (54.9~eV) is just
beyond the \HeII\ edge
where normal stars emit little flux and hence, the
origin of such high-excitation line in starbursts has been unclear. Different
excitation mechanisms such as weak AGNs, super-hot stars (e.g. WR
stars), planetary nebulae and ionising shocks related to the starburst
activity have been proposed by \cite{lutz98}. These authors favour
ionising shocks as the most likely explanation for the presence of
\OIV\ emission in starburst galaxies. \cite{schaerer99}, on the other
hand, propose that the emission of \OIV\ 25.9~\micron\ observed with
ISO in the two dwarf galaxies NGC\,5253 and II\,Zw\,40 is due to the
presence of hot WR stars.
Independently of the origin of the \OIV\ emission,
the mere presence of such high energy source(s) in NGC\,5253
calls for an examination of their possible effects on the
photoionisation models presented above and on the robustness
of the results. We shall briefly address this now.

\begin{figure}[!ht]
   \centering \includegraphics[width=8.8cm]{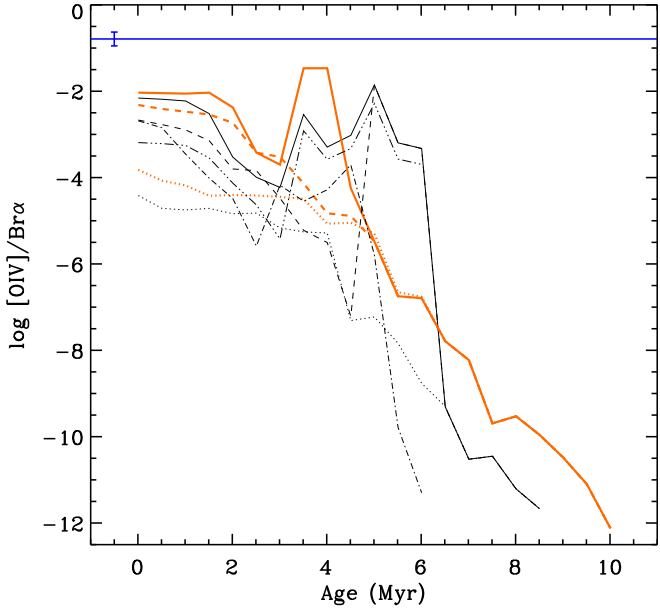}
   \caption{Variation of \OIV/Br$\alpha$ as a function of starburst
     age. The observed ratio is indicated by an horizontal line and
     its uncertainty by an error bar on the left side. Models are
     plotted using the same code as in Fig.~\ref{fig:cloudy_1}.}
        \label{fig:oiv}
   \end{figure}

First we assume that the  \OIV\ 25.9~\micron\ emission observed
in the ISO-SWS aperture originates from the cluster C2.
Figure \ref{fig:oiv} shows the observed and predicted \OIV/Br$\alpha$
line ratio from the starburst models discussed above (Figs.\ \ref{fig:cloudy_1}
and \ref{fig:cloudy_2}).
Manifestly, the predicted strength of the \OIV\ line is much lower than 
observed.
In the best case (the model with \mup=100\,\msun\ and $Z=0.004$),
\OIV/Br$\alpha$ reaches few percent ($\sim 3$\%) between $\sim$ 3 to 5 Myr 
when WR stars
are present.
In contrast to the models of \cite{schaerer99}, the discrepancy between the
observed and predicted  \OIV\ emission found here is due to
{\em 1)} softer stellar atmosphere models (mostly WR spectra including
line blanketing; \citeauthor{smith02} \citeyear{smith02}), and {\em 2)}
a reliable constraint on the ionisation parameter.

{\em Can the models be reconciled with the observed \OIV\ line flux?}

In principle this could be possible by increasing the ionisation parameter $U$
-- e.g.\ by adopting a smaller inner radius and/or a density gradient and/or 
internal dust
and/or a top-heavy IMF (see Sect.~\ref{sect:models}).
However, as discussed previously, such an increase of $U$
would cause a larger discrepancy between the models with \mup=100\,\msun\
and the other mid-IR lines. In particular,
the \ArIII\ line flux, which is barely reproduced by the fiducial models with
\mup=100\,\msun\ would be largely underpredicted.

A more promising way to circumvent the shortfall of the observed \OIV\
line flux is by considering that the adopted spectral energy
distribution of the cluster close to and beyond the \HeII\ edge
is not well defined.
This could be justified for several reasons.
First, X-emission in early type stars can alter their ionising spectra
at high energies
\citep[see e.g.][ and discussion therein]{pauldrach01,schaerer97,macfarlane94}.
Second, other sources of high energy photons as binaries or shocks responsible
of soft X-ray emission could be present in C2  \citep{summers04}.
In this situation, the ionising spectrum could be increased at high
energy such as to enhance the \OIV\ emission without altering
the ionisation structure (and hence the emission line strengths) of the
other observed ions whose ionisation potentials are lower ($IP \sim$
30--40 eV).
Indeed, after several ad-hoc modifications of the library SEDs beyond $\sim
45$~eV, we have verified that a spectral energy distribution with
$\log Q_2/Q_0 \sim -1.8$\footnote{Here $Q_0$ and $Q_2$ stand for the
ionising flux capable to ionise H and He$^+$ respectively.}
can reproduce both the observed \OIV\ line
flux and the other constraints.
This corresponds to a harder SED than typically expected for starbursts
 \citep[cf.][]{smith01}.
For instance, the hardest {\sc Starburst99} model used above (a $Z=0.004$ 
burst at 4~Myr) has
$\log Q_2/Q_0 = -2.7$, whereas in other models one typically has
$\log Q_2/Q_0 \sim$ --5 to --6.
In fact, if the X-ray luminosity determined for region \#19 by
\cite{summers04} is attributed to C2, one has $L_X/L_{\rm bol} \sim 10^{-4.4}$,
considerably higher than the ``normal'' X-ray luminosity of early type
stars \citep{chlebowski89}.
Invoking a harder SED
than predicted by standard synthesis models would therefore seem justified.
Finally the postulated hardness of $\log Q_2/Q_0 \sim -1.8$
would correspond to an optical line ratio of $I(HeII \lambda4686)/I(H\beta)
\sim 3\%$, a hardness not unusual in metal-poor \HII\ regions and
Wolf-Rayet galaxies \citep[cf.][]{guseva00}.

From the above we see that is possible to construct models
able to reconcile the observed ISO \OIV\ 25.9~\micron\ emission
with the other observables for the cluster C2 without altering our conclusions
on the age, IMF etc. Furthermore, the required harder ionising spectrum
does not seem physically unplausible.
On the other hand, it is not excluded that part or all of the \OIV\
line flux is not related to C2 and comes from another object included
in the large ISO-SWS aperture.
In this case, our observations and modelling require, for consistency, that
this source shows a rather faint  \ArIII\ 9.0 \micron\ emission and
a \SIV\ 10.5 \micron\ comparable or less than C2.
Mid-infrared observations with Spitzer should be
able to test these predictions.

\section{Discussion}
\label{sect:discussion}

\subsection{Importance of high spatial resolution observations}

We have shown the importance of high spatial resolution observations
of extra-galactic star clusters. Only such observations provide
accurate measurements of the line fluxes emitted by the cluster itself
and avoid contamination by other clusters, the local ISM, etc.  For
instance, in the case of NGC\,5253~C2, while most of the
high-excitation \SIV\ and \NeIII\ line fluxes measured by the large
ISO aperture is emitted by the \HII\ region associated to the cluster,
other sources largely contribute to the low-excitation \ArII, \SIII\
and \NeII\ line fluxes.
For example, the remaining presumably ``diffuse'' \NeII\ emission
in the large ISO aperture amounts to $\sim$ 80\% of the total emission.

This has implications not only in the interpretation of line
fluxes in term of the properties (age, IMF, etc.) of the embedded cluster
\citep[e.g.][]{crowther99,thornley00,rigby04},
but also on the modelling of the SED \citep[e.g.][]{vanzi04},
where contamination by background radiation
from the underlying galaxy and other sources might lead to erroneous
conclusions.

High spatial resolution observations also provide important
constraints on the nebular geometry (size, electron density,
etc.). The interpretation of nebular lines in terms of properties of
the ionising cluster depends on photoionisation models which have as
free parameters the SED of the cluster ($M_{\rm up}$ and age), the
local abundance and the ionisation parameter $U$. In particular, the
parameter $U$, which depends on the geometry, is generally strongly
underconstrained, leading to wrong conclusions. Along these lines,
we have shown how two SEDs with different $M_{\rm up}$ can lead to
similar conclusions when two different geometries are considered (see
Sect.~\ref{sect:u}). The same holds when the local abundance is not properly
taken into account (cf. Sect.~\ref{sect:solar}).

We have thus demonstrated that the spatial scale of the observations greatly
determine the mid-IR appearance of NGC\,5253. Similar aperture effects
affecting both the line and PAH emission fluxes have been observed by
\cite{siebenmorgen04} in luminous IR galaxies.
However, it is likely that ``global'' mid-IR (ISO or even the smaller Spitzer
aperture) diagnostics give to a large extent correct
answers concerning AGN/starburst diagnostics. This is supported
by the comparison with optical diagnostics, which mostly agree
with the mid-IR ones \citep[cf.][]{genzel98,laurent00,peeters04}.
However, more detailed and subtle diagnostics (IMF, age, etc.)
are probably more prone to aperture effects.

\subsection{Constraints on the age and upper mass cutoff}

It has been claimed that the IMF upper mass cutoff is not well
defined, but is given by the stochastic nature of the IMF
\citep[cf.][]{elmegreen97}. However, NGC\,5253~C2 is massive enough
(see Table~\ref{table:clusters}) to be populated with stars up to
\mup\ \citep[cf.][]{weidner04}.

For NGC\,5253~C2, strong constraints on the geometry (i.e. Lyman
photon rate, density, filling factor and inner/outer radius of the
nebula) based on high spatial observations in the near-IR (HST and
Keck), mid-IR (TIMMI2) and radio (VLA) lead to an ionisation parameter
$\log U \geq -0.5$~dex.  This restriction on the parameter $U$ allows
us to conclude that a cluster with an upper mass cutoff of 100\,\msun\
can only reproduce the observed line fluxes if its age is about
5--6~Myr, above the upper limit set by the non presence of supernovae
($\sim 4$~Myr).  A cluster younger than about 4~Myr can fit the
observations if its $M_{\rm up} \lesssim 50\,M_{\sun}$.  A young
cluster of 1~Myr would only be consistent with an even lower upper
mass cutoff (around 30\,M$_{\sun}$).  This result is practically
independent on the power law index of the IMF
(cf. Sect.~\ref{sect:alpha}), the internal dust content
(cf. Sect.~\ref{sect:dust}) and the density profile
(cf. Sect.~\ref{sect:dprofile}).

The photoionisation models allow higher values of the upper mass
cutoff for ages $< 4$~Myr only in the case of an \HII\ region much
larger than what the radio continuum images imply
(cf. Sect.~\ref{sect:u}) or when a solar metallicity, for which there
is no indication, is considered for the cluster
(cf. Sect.~\ref{sect:solar}).

In short, we are left schematically with two possible solutions due to
an age degeneracy: {\em 1)} a young ($<$ 4 Myr) cluster with a clear
evidence for a ``non-standard'', low, upper mass cutoff of the IMF, or
{\em 2)} a cluster of $\sim$ 5--6 Myr with a ``standard'' high  upper
mass cutoff.
The existing age constraints have been described above (Sect.\
\ref{sect:constrains}).  Let us now discuss these two cases and their
implications.

\subsubsection{A young cluster deficient in massive stars?}

A young age ($\la$ 3--4 Myr) would agree with the lack of supernovae
signatures in C2, but it would imply a ``non-standard'' IMF with a low
upper mass cutoff ($\mup \la 50 \msun$).  Indeed, so far there is a
general consensus from various studies of individual stars in clusters
that stars of at least $\sim$ 100\,\msun\ are formed and that most
likely \mup\ $\sim$ 120--150\,\msun\ quite independently of
metallicity and other ``environmental'' conditions
(e.g. \citeauthor{massey98} \citeyear{massey98}, \citeauthor{figer99}
\citeyear{figer99}, \citeauthor{kroupa02} \citeyear{kroupa02},
\citeauthor{weidner04} \citeyear{weidner04} and reviews in
\citeauthor{gilmore98} \citeyear{gilmore98} and
\citeauthor{corbelli04} \citeyear{corbelli04}).  $\mup > 50 \msun$ has
also been found from integrated light studies of (giant)
extra-galactic \HII\ regions, \HII\ galaxies and alike objects at
sub-solar metallicities (cf. reviews of \citeauthor{leitherer98}
\citeyear{leitherer98} and \citeauthor{schaerer03}
\citeyear{schaerer03}).  Even at high metallicity, recent direct
studies of the stellar content indicate a high/normal value of \mup\
\citep{pindao02}.  Claims for a low upper mass cutoff of the IMF have
been made repeatedly
\citep[e.g.][]{goldader97,bresolin99,coziol01,rigby04}.  However,
these claims are essentially based on less direct diagnostics
involving nebular line observations, large apertures and/or objects
with complex structures which render their interpretation more
difficult.
Here we find, for the first time, an indication for a
``non-standard'', low upper mass cutoff of the IMF for an individual
massive cluster.  If true, this implies that the IMF is different and
not fully populated up to the largest masses in some regions/objects.
What is responsible for this deviation remains, however, unknown as
apparently similar clusters with a normal IMF exist (cf.\ list in
Table \ref{table:clusters}).  Is it related to the cluster mass, to
its strong gravitational potential, or dependent on other properties?
Do dynamical processes and interactions play a role in establishing
the IMF in such a case?  A better understanding of the IMF and its
origin are needed before these questions can be answered.

What are the implications from a massive star deficiency?
This obviously depends on the physical quantity of interest
and on how general such a case is. Studies of additional hidden
SSCs will obviously be of interest to examine if our result is
also found for other objects.
For example, we simply note that
the bolometric luminosity emitted by a massive young ($\la$ 4--5~Myr)
cluster with $\mup = 30$\,\msun\ is $\sim$ 1 mag smaller per unit stellar
mass than for a normal $\mup =$ 100\,\msun. In other words the
contribution of such a hidden cluster to the dust heating and IR flux
is reduced by the same amount.

\subsubsection{An ``old'' hidden compact cluster?}

The solution of an ``old'' age of $\sim$ 5--6~Myr raises several
question.  First, how is it possible to ``contain'' and hide such a
compact cluster so  long?  With a much higher electron density ($\sim
5\times10^4$~cm$^{-3}$) than what it is commonly found in typical
giant extra-galactic \HII\ regions ($\sim 10^2-10^3$~cm$^{-3}$),
NGC\,5253~C2 should be overpressured compared to the surrounding
ionised gas. At such high pressure, the region would expand rapidly to
reestablish pressure equilibrium. Following \cite{wood89}, we find
that with an age of 5--6~Myr and under no other additional external
pressure, C2 would have expanded until a size of 22--24~pc in
diameter. However, this size is more than 10 times what it is observed
at radio wavelengths (1.6 by 0.6~pc, cf. Sect.~\ref{sect:constrains}).
The most simple explanation seems that the cluster mass is large
enough to allow its own gravitational confinement (self-gravity), as
suggested by  \cite{turner03} and \cite{tan04}.

While several other compact hidden clusters have been known for
several years \citep{kobulnicky99,neff00,beck02} they are generally
thought to be very young with ages ($\ll$ 1--4 Myr) mainly based on
compactness arguments. In some cases also, compact thermal radio
sources seem to show optical counterparts providing age estimates of
$\sim$ 0--4 Myr \citep{whitmore02}.
Here, however, our analysis indicates for the first time that such
hidden clusters could live longer.  The ``average'' fraction of the
lifetime massive stars spend buried in their parent molecular clouds
can also be estimated by comparing the number of ``hidden'' O stars to
the total observed in the galaxy. The hidden cluster C2 is powered by
$\sim 2\times10^{52}$ Lyman continuum photons per second, which
implies about 2000 O7\,V equivalent stars. Over a 30\arcsec\ aperture,
the total (extinction-corrected) Br$\gamma$ line flux is $\sim
5\times10^{-20}$~W~cm$^{-2}$, which is equivalent to
$5\times10^{52}$~s$^{-1}$ and 5000 O7\,V equivalent stars. Based on
these estimates, O stars in NGC\,5253 might spend approximately 40\%
of their lives in the embedded phase. This percentage is much larger
than the 10\%--20\% estimated for Galactic ultracompact \HII\ regions
\citep[cf.][]{wood89b} and the 15\% estimated for young ultradense
\HII\ regions in the young starburst galaxy He\,2--10 by
\cite{kobulnicky99}.

If this scenario is true, it remains to be understood what determines
the duration of the ``hidden'' phase of SSCs allowing for significant
variations of the latter. Is it the cluster mass, concentration, its
gravitational potential, external effects or others? The only clue we
have so far is that the cluster C2 analysed here is among the most
massive and compact ones (see Table~\ref{table:clusters}).  Future
detailed studies of young massive clusters should be able to able to
shed more light into these questions.

On the other hand, the ``old'' solution of $\sim$ 5--6 Myr is somewhat
puzzling given the apparent absence of supernovae in this region of
NGC 5253 (cf. \ref{sect:constrains}).  Indeed, given the high
luminosity/Lyman continuum production of C2 (see Table
\ref{table:clusters}) one expects a supernova rate of at least 100
SN/Myr after approximately 3 Myr \citep[e.g.][]{cervino00} or in other
words, at least a total of several hundred past SN explosions at an
age of  $\sim$ 5--6 Myr from stars with initial masses $M \ga$ 35--40
\msun.  However, it seems difficult to understand how such a large
number of supernovae could be ``hidden'' without giving raise to
non-thermal radio emission especially in such a high density region
\citep[cf.][]{chevalier01}.



\begin{table*}[!ht]
\caption{Properties of massive clusters. Given are the total cluster
  mass, radius, mass density in stars, age, luminosity and rate of
  Lyman photons.}
  \label{table:clusters}
  \begin{center}
    \leavevmode
    \begin{tabular}[h]{lccccccc}
      \hline \hline \\[-7pt]
     \multicolumn{1}{c}{Cluster} &
     \multicolumn{1}{c}{$\log M/M_{\sun}$} &
     \multicolumn{1}{c}{Radius (pc)} &
     \multicolumn{1}{c}{$\log \rho (M_{\sun} {\rm pc}^{-3})$} &
     \multicolumn{1}{c}{Age (Myr)} &
     \multicolumn{1}{c}{$\log L/L_{\sun}$} &
     \multicolumn{1}{c}{$\log Q_0 ({\rm s}^{-1})$} &
     \multicolumn{1}{c}{Ref.}
   \\[5pt] \hline \\[-7pt]

NGC\,5253~C2 & $<6.5^\diamond$ & 0.8 & $<6.2$ & $<6$    & 9.0    & 52.3 & 1, 2, 3, 4\\
Quintuplet   & 3.8$^\star$ & 1.0  & 3.2 & 3--6   & 7.5    & 50.9 & 5\\
Arches       & 4.3$^\star$ & 0.19 & 5.8 & 1--2   & 8.0    & 51.0 & 5\\
Gal. Center  & 4.0$^\star$ & 0.23 & 5.6 & 3--7   & 7.3    & 50.5 & 5\\
NGC\,3603    & 3.7$^\star$ & 0.23 & 5.0 & 2.5    & 7.3    & 51.1 & 5\\
R136         & 4.5$^\star$ & 1.6  & 3.3 & $<1-2$ & $>7.6$ & 51.9 & 5\\
MGG--11$^\triangle$   & 5.5$^\flat$ & 1.2  & 4.7 & 7--12  & ...    & ...  & 6\\
$[$W99$]$2$^\natural$ & 6.3$^\flat$ & 4.5  & ... & 7      & ...    & ...  & 7\\[5pt]

\hline

    \end{tabular}
  \end{center}
$^\diamond$ If due to virialised motions, the Br$\gamma$ line width measured
by \cite{turner03} indicates a cluster mass of the order of $10^6$\,\msun\ for a
Salpeter IMF slope and a lower mass cutoff of $\sim 1$\,\msun\ or less.
$^\star$ Total cluster mass in all stars extrapolated down to a
lower mass cutoff of 1\,\msun, assuming a Salpeter IMF slope and an upper mass
cutoff of 120\,\msun.
$^\flat$ Kinematic mass.
$^\triangle$ Star cluster in M82.
$^\natural$ Star cluster in NGC\,4038/4039.
REFERENCES:
(1) This work; 
(2) \cite{turner04}; 
(3) \cite{vanzi04}; 
(4) \cite{alonso04}; 
(5) \cite{figer99b};
(6) \cite{mccrady03};
(7) \cite{mengel02}.
\end{table*}

\subsection{Comparison with previous mid-IR constraints on the IMF}

In the case of NGC\,5253 we are facing an apparent ``contradiction''.
The ISO observations show that NGC\,5253 is, with II\,Zw\,40, the
object with the highest excitation within the ISO extra-galactic
sample (as measured by \NeIII/\NeII).  For this reason this object is
considered as ``unproblematic'', i.e.\ compatible with a normal
Salpeter IMF extending to at least $\sim$ 40--60 \msun\
\citep{rigby04} in contrast to the majority of the starbursts observed
by ISO.  In addition, our observations now indicate an {\em even
higher excitation} for the SSC C2.  {\em Why then do we here face this
difficulty of reconciling the observations with a ``normal'', high
value of \mup ?}  The answer to this apparent contradiction is simple:
the difficulty arises from the fact that a proper geometrical
constraint is used here for the first time implying a fairly large
ionisation parameter.  Had we neglected the information on the spatial
extent (or better said the compactness) of the emission associated
with C2, we would easily be able to reconcile the observations with a
standard value of \mup\ for any burst age.  This case illustrates the
potentially severe limitations inherent in the traditional
interpretations of fine structure line ratios when the source
distribution and geometry are unknown.  However, from the single case
study presented here it is not possible to generalise the conclusions.

\subsection{Comparison with other massive clusters}

A stellar cluster with \mup = 100\,\msun\ and an age of 6~Myr would have
a total cluster mass $M_{\rm cluster} \sim 34\times10^5$\,\msun\ and a 
total number
of stars $N_{\rm stars}= 9\times10^5$ for a Salpeter IMF slope extending down
to 1\,\msun.
A stellar cluster with \mup = 50\,\msun\ and an age of 4~Myr
would have $M_{\rm cluster} \sim 9\times10^5~M_{\sun}$ and
$N_{\rm stars} \sim 2 \times10^5$.  Finally, a stellar cluster with
$M_{\rm up}= 30\,M_{\sun}$ and 1~Myr
would have $M_{\rm cluster}=4\times10^5~M_{\sun}$ and $N_{\rm stars}=
1\times10^5$.

Table~\ref{table:clusters} compares NGC\,5253~C2 and other massive
clusters in mass, size, density, age, luminosity and Lyman continuum
flux. NGC\,5253~C2 is most comparable in mass to MGG--11 and $[$W99$]$2,
but it is
younger. In terms of mass density, NGC\,5253~C2 is comparable to
the Galactic Centre and Arches clusters, but it has a higher
luminosity and produces more ionising photons.

Regarding the upper mass cutoff,
studies of R136, the ionising
cluster of 30\,Doradus, find that the most probable upper mass limit in
the Large Magellanic Cloud is about 130\,M$_{\sun}$ \citep[cf.][]{selman99}.
Towards the Arches cluster, \cite{figer99} find that there
more than 10 stars with masses larger than 120\,M$_{\sun}$.
If true, the case of a low upper mass cutoff in NGC\,5253~C2 would
greatly differ with the high values found in these massive clusters.

\subsection{Host of an intermediate mass black hole?}
\label{sect:bh}

Recent dynamical simulations by \cite{portegies04} show that massive
stars in young dense clusters can sink rapidly to the centre of the
cluster to participate in a runaway collision and produce a star of
800--3000\,$M_{\sun}$ which ultimately collapses to a black hole of
intermediate mass (see also \citeauthor{gurkan04} \citeyear{gurkan04}
and references therein).

The conditions for the formation of a black
hole inside a dense cluster depend on two distinct parameters, the
concentration parameter ($c$) and the dynamical friction timescale
($t_{\rm df}$).
The parameter $c$ is defined as $c \equiv \log (R_{\rm t}
/ R_{\rm c})$, where $R_{\rm t}$ and $R_{\rm c}$ represent the tidal
radius and the core radius, respectively; $t_{\rm df}$ is the time
taken for a star of a certain
mass to sink to the cluster centre from a circular orbit at initial
distance $R\ggg R_{\rm c}$.
According to the simulations by \cite{portegies04}, the
requirement for the formation of such a black hole is that the cluster
is born with $c \gtrsim 2$ and that $t_{\rm df} \lesssim 4$~Myr.
The 4~Myr upper limit for $t_{\rm df}$ is
such that massive stars can easily reach the centre of the cluster
before exploding as supernovae.These authors' simulations indicate
that in the cluster MGG--11, located in M82 and associated with an
ultraluminous X-ray source, a runaway star could form on the necessary
timescale. In the case of NGC\,5253~C2, we find $t_{\rm df} \sim
2.5$~Myr, calculated for a star of 50\,$M_{\sun}$ starting at a radius
of 0.8~pc. As stated by \cite{portegies04}, less massive stars undergo
weaker friction and thus must start at smaller radii in order to reach
the cluster centre on a similar timescale. Hence, an intermediate mass
black hole might form through runaway collisions in the core of
NGC\,5253~C2.

Recent Chandra and XMM-Newton observations of NGC\,5253
\citep{summers04} have found a X-ray point source at the peak of the
radio emission associated with C2 (their source \#19). However, this
X-ray source is very soft, with very few counts above 1.0 keV and
virtually none above 2.0 keV. The authors suggest that this X-ray
emission is associated with the emission from superbubles surrounding
the inner cluster rather than with an individual harder object such as a
black hole. Besides, they find that the X-ray emission coincident with
C2 seems to be extended compared to the Chandra PSF, reinforcing the
idea that the emission is associated with the cluster. Hence, although
it seems that X-ray
observations show no evidence for the presence of a black hole within C2,
it is a possibility that deserves certain consideration.

\section{Conclusions}
\label{sect:conclusions}

We have presented the $N$-band (8--13~\micron) spectrum of the hidden
compact radio super-star cluster in NGC\,5253, C2, obtained with
TIMMI2 on the ESO 3.6\,m telescope. The spectrum is characterised by a
rising continuum due to warm dust, a silicate absorption and a strong
\SIV\ line at 10.5~\micron. Weaker lines of \ArIII\ at 9.0~\micron\
and \NeII\ at 12.8~\micron\ are also present.  The continuum can be
modeled by an optically thick emission from hot ($T_{\rm d}=253 \pm
1$~K) dust emission extinguished by a cold foreground dust screen and
a silicate absorption feature with $A_{\rm sil} = 0.73 \pm 0.05$ mag.

It has been demonstrated that the spatial scale of the observations
greatly determine the mid-IR appearance of NGC\,5253. In particular,
most of the high-excitation \ArIII, \SIV\ and \NeIII\ line fluxes
measured by the large ISO aperture are emitted by C2, while other
sources largely contribute to the low-excitation \ArII, \SIII\ and
\NeII\ line fluxes. This has important implications in the
interpretation of line fluxes in terms of the properties (age, IMF
etc.) of the embedded cluster. For instance, the present data indicate
an even higher excitation for C2 compared to the observations through
the large ISO/SWS aperture.

We have computed sets of nebular models with the photoionisation code
CLOUDY using the evolutionary synthesis code {\sc Starburst99} to
model the integrated properties of the stellar cluster. It has been
shown that the interpretation of the nebular lines in terms of
properties of the ionising cluster depends largely on the local
abundance and the ionisation parameter $U$. The detailed dependence of
the mid-IR lines on other parameters such as the cluster age, upper
mass cutoff and power law index of the IMF, as well as the presence of
internal dust and the density structure is largely discussed. In the
case of the SSC C2, high spatial resolution observations at different
wavelengths -- near-IR (HST and Keck), mid-IR (TIMMI2) and radio (VLA)
-- have allowed us to strongly constrain the geometry of the region
(i.e. Lyman photon rate, density, filling factor and inner/outer
radius), leading to an ionisation parameter $\log U \ge
-0.5$~dex. This constraint on $U$ lead to two possible solutions for
the age and upper mass cutoff of C2 when comparing the observed line
fluxes with the predicted ones: {\em 1)} a young ($< 4$~Myr) cluster
with a "non-standard" IMF with a low upper mass cutoff \mup $<
50$\,\msun, and {\em 2)} a cluster of $\sim 5-6$~Myr with a standard
high upper mass cutoff (\mup $\sim 100$\,\msun). The photoionisation
models allow higher values of \mup\ for ages $< 4$~Myr only in the
case of an \HII\ region much larger than what the radio continuum
imply or when a solar metallicity, for which there is no indication,
is considered for the cluster.  A young age ($<$ 4~Myr) would agree
with the lack of supernovae signatures in C2 and in case of being
confirmed, would be the first indication for a ``non-standard'', low
upper mass cutoff of the IMF for an individual massive cluster. An
older age of $\sim$ 5--6~Myr would imply that it is possible to
``contain'' and hide such a compact cluster for a longer time that
what it is generally thought.  Arguments in favour and against these
two scenarios are presented. Based on the available information, we
are not able to favour one solution over the other. A better
understanding of the signature of supernovae from massive star
progenitors in compact and dense environments, as well as radio
observations at even higher spatial resolution, would permit to
exclude or not the ages above 4~Myr.
Besides, mid-IR observations at even higher spatial resolution (which can be performed e.g. in the near future with VISIR on the VLT), would allow determining the exact size of C2 at these wavelengths and resolving its ionisation structure. In particular, resolving the ionsation structure of C2 would be useful to verify if the lines we are using for our photoionisation analysis come from the same region or if there is any contamination by the close presence of the optical stellar cluster C1.
Observations of massive stellar
clusters with similar characteristics will be as well essential to see
if the same result is found.

We have addressed as well the origin of the \OIV\ 25.9~\micron\
emission measured by ISO assuming that all of it originates from the
SSC C2.  The strength of the \OIV\ line predicted by the
photoionisation models is much lower than observed. However, it has
been shown that it is possible to construct models able to reconcile
the observed ISO \OIV\ emission with the other mid-IR lines by
modifying the adopted SED of the cluster close to and beyond the
\HeII\ edge. A SED with $\log Q_2/Q_1 \sim -1.8$ can reproduce all the
observables. This hardness would correspond to an optical line ratio
of $I(HeII\lambda4686)/I(H\beta) \sim 3\%$, not unusual in metal-poor
\HII\ regions and Wolf-Rayet galaxies.

Finally, based on dynamical simulations, we consider the possibility
that an intermediate mass black hole might form through runaway
collisions in the core of C2. However, recent X-ray observations show
that the X-ray source associated with C2 is very soft and it is most
probably associated with the emission from superbubles surrounding the
inner cluster.

\begin{acknowledgements}
We have benefited from numerous discussions with and suggestions from a
number of colleagues. Among them we would like to thank
Grazyna Stasi\'nska, Claus Leitherer and Berhard Brandl in particular.
We also thank Almudena Alonso-Herrero for communicating results
prior to publication. Finally, we want to thank Els Peeters for
providing us with fully reduced ISO/SWS spectra and Ilaria Pascucci for 
letting us use
her IDL routine to fit the mid-IR dust continuum.
This work was supported by the Swiss National Foundation, the
French {\it Centre National de la Recherche Scientifique},
and CEA.

\end{acknowledgements}

\appendix

\section{The 18~\micron\ flux of NGC\,5253~C2}

Besides the $N$-band spectrum, we also obtain an image of NGC\,5253~C2
in the $Q1$ band (50\% band cuts are 17.35 and 18.15~\micron). The
object appears unresolved, with a FWHM of $2\farcs4\pm0\farcs1$. The
flux was calibrated using the standard star HD\,123139. We obtain a
total flux of $4.53\pm0.25$~Jy. This value is $\sim 1.6$ times larger
than the 18.7~\micron\ flux reported by \cite{gorjian01}; $2.9\pm0.3$
Jy. However, according to Gorjian (private communication), their low
18.7~\micron\ flux could be due to a too small chopping throw,
resulting in an incorrect zero-level, or to a complex airmass
correction.

\bibliographystyle{aa}

\end{document}